\documentclass[10pt,twocolumn,letterpaper]{article}

\usepackage{cvpr}
\usepackage{times}
\usepackage{epsfig}
\usepackage{graphicx}
\usepackage{amsmath}
\usepackage{amssymb}

\usepackage{multirow}
\usepackage{xcolor}
\usepackage{float}
\usepackage{subfigure}
\usepackage{balance}

\usepackage[pagebackref=true,breaklinks=true,letterpaper=true,colorlinks,bookmarks=false]{hyperref}

\cvprfinalcopy 

\def\cvprPaperID{963} 

\ifcvprfinal\fi
\begin{document}

\definecolor{HKKUO}{rgb}{0.36, 0.54, 0.66}
\definecolor{YS}{rgb}{0.0, 0.67, 0.28}
\definecolor{roy}{rgb}{0.45, 0.31, 0.59}
\definecolor{Check}{rgb}{0.82, 0.1, 0.26}
\definecolor{Ulia}{rgb}{0.7, 0.45, 0.2}
\definecolor{YM}{rgb}{0.8, 0.3, 0.6}

\title{Unified Dynamic Convolutional Network for Super-Resolution with \\ Variational Degradations}

\author{
Yu-Syuan Xu, Shou-Yao Roy Tseng, Yu Tseng, Hsien-Kai Kuo, and Yi-Min Tsai\\
MediaTek Inc., Hsinchu, Taiwan\\
{\tt\small \{Yu-Syuan.Xu, Roy.Tseng, Ulia.Tseng, Hsienkai.Kuo, Yi-Min.Tsai\}@mediatek.com}
}

\maketitle
\thispagestyle{empty}

\begin{abstract}
Deep Convolutional Neural Networks (CNNs) have achieved remarkable results on Single Image Super-Resolution (SISR). Despite considering only a single degradation, recent studies also include multiple degrading effects to better reflect real-world cases. However, most of the works assume a fixed combination of degrading effects, or even train an individual network for different combinations. Instead, a more practical approach is to train a single network for wide-ranging and variational degradations. To fulfill this requirement, this paper proposes a unified network to accommodate the variations from inter-image (cross-image variations) and intra-image (spatial variations). Different from the existing works, we incorporate dynamic convolution which is a far more flexible alternative to handle different variations. In SISR with non-blind setting, our Unified Dynamic Convolutional Network for Variational Degradations (UDVD) is evaluated on both synthetic and real images with an extensive set of variations. The qualitative results demonstrate the effectiveness of UDVD over various existing works. Extensive experiments show that our UDVD achieves favorable or comparable performance on both synthetic and real images.
\end{abstract}

\section{Introduction}
Single image super-resolution (SISR) has posed a great challenge of image quality in the recent advance of computer vision. The goal of SISR is to reconstruct a High-Resolution (HR) image from a Low-Resolution (LR) image. This inverse property makes it a highly ill-posed problem. Deep Convolutional Neural Networks (CNNs)~\cite{SRCNN,VDSR,DRCN,DRRN,MemNet,SRGAN,ESRGAN,MSLapSRN,FSRCNN,ESPCNN,EDSR,RDN,DBPN,RCAN} have been widely adopted to solve SISR problem, and achieve significant success. 
Nevertheless, most of the methods assume a single fixed combination of degrading effects, \eg, blurring and bicubicly downsampling. Such assumption limits their capability to handle practical cases with multiple degradations.

\begin{figure}[t]
\centering
\footnotesize
\begin{minipage}{0.195\textwidth}
\boldmath
\centerline{\includegraphics[width=\textwidth]{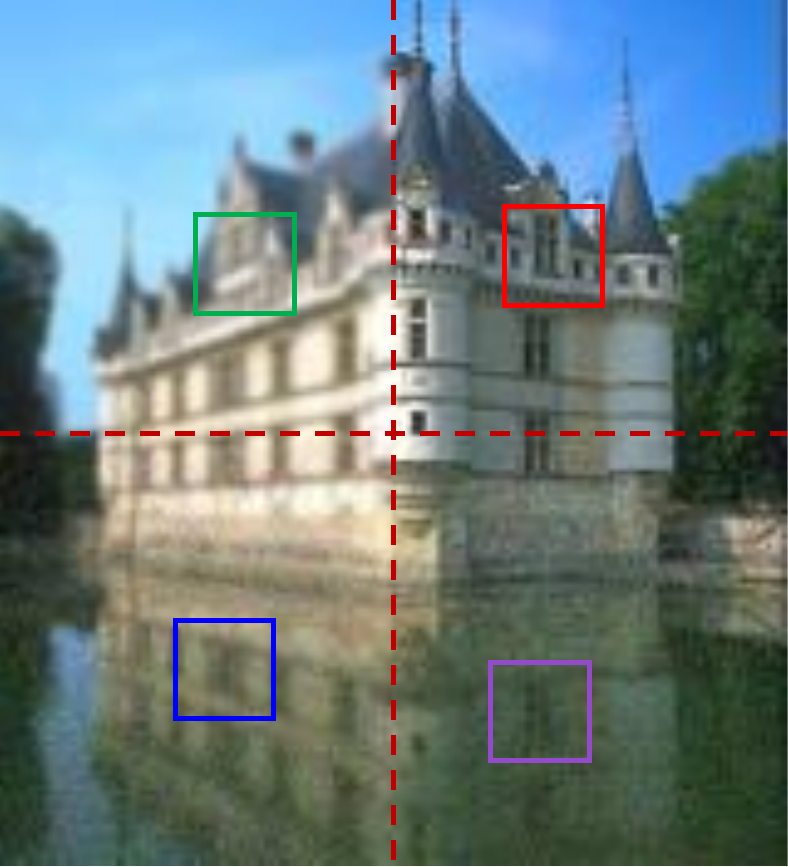}}
\vspace{-110pt}
\centerline{\textcolor{white}{\textbf{$\varepsilon$=1.6, $\sigma$=0 \hspace{10pt} $\varepsilon$=0.2,$\sigma$=0}}}
\vspace{88pt}
\centerline{\textcolor{white}{\textbf{$\varepsilon$=1.6, $\sigma$=10 \hspace{6pt} $\varepsilon$=0.2, $\sigma$=10}}}
\centerline{Input Image}
\end{minipage}
\begin{minipage}{0.062\textwidth}
\vspace{-2pt}
\centerline{\includegraphics[width=\textwidth]{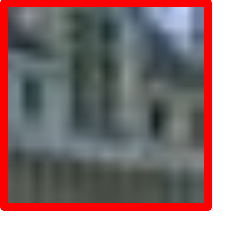}}
\vspace{-4.5pt}
\centerline{\includegraphics[width=\textwidth]{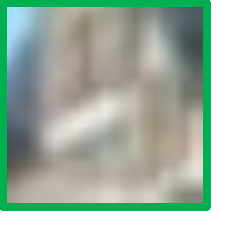}}
\vspace{-4.5pt}
\centerline{\includegraphics[width=\textwidth]{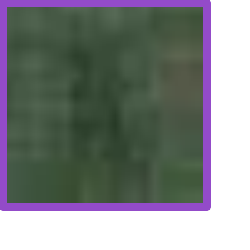}}
\vspace{-4.5pt}
\centerline{\includegraphics[width=\textwidth]{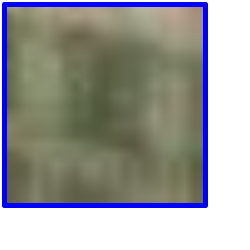}}
\vspace{-8pt}
\centerline{\hspace{-4pt}LR}
\end{minipage}
\hspace{-7.7pt}
\begin{minipage}{0.062\textwidth}
\vspace{7pt}
\centerline{\includegraphics[width=\textwidth]{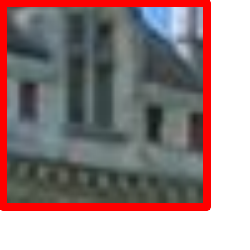}}
\vspace{-4.5pt}
\centerline{\includegraphics[width=\textwidth]{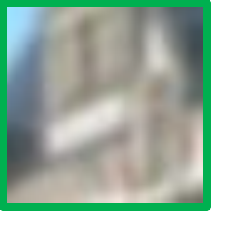}}
\vspace{-4.5pt}
\centerline{\includegraphics[width=\textwidth]{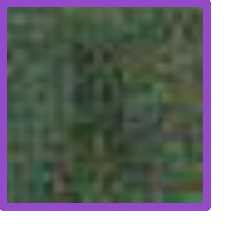}}
\vspace{-4.5pt}
\centerline{\includegraphics[width=\textwidth]{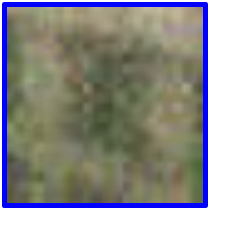}}
\vspace{-8pt}
\centerline{\hspace{-4pt}RCAN}
\vspace{-2pt}
\centerline{\hspace{-4pt}\cite{RCAN}}
\end{minipage}
\hspace{-7.7pt}
\begin{minipage}{0.062\textwidth}
\vspace{7pt}
\centerline{\includegraphics[width=\textwidth]{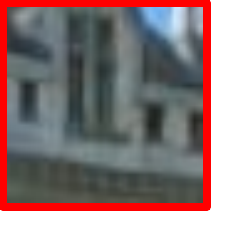}}
\vspace{-4.5pt}
\centerline{\includegraphics[width=\textwidth]{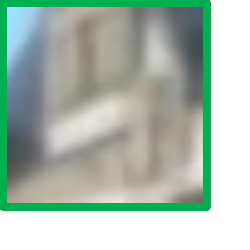}}
\vspace{-4.5pt}
\centerline{\includegraphics[width=\textwidth]{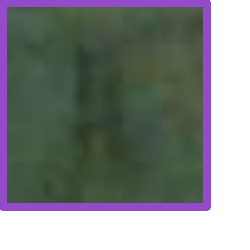}}
\vspace{-4.5pt}
\centerline{\includegraphics[width=\textwidth]{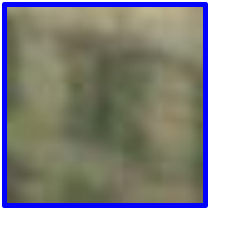}}
\vspace{-8pt}
\centerline{\hspace{-4pt}ZSSR}
\vspace{-2pt}
\centerline{\hspace{-4pt}\cite{ZSSR}}
\end{minipage}
\hspace{-7.7pt}
\begin{minipage}{0.062\textwidth}
\vspace{7pt}
\centerline{\includegraphics[width=\textwidth]{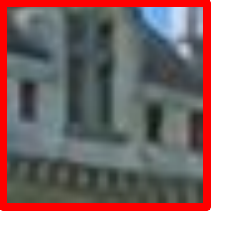}}
\vspace{-4.5pt}
\centerline{\includegraphics[width=\textwidth]{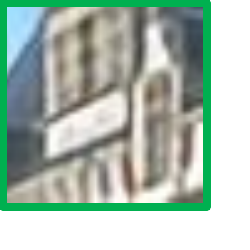}}
\vspace{-4.5pt}
\centerline{\includegraphics[width=\textwidth]{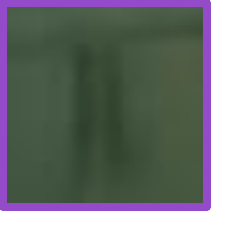}}
\vspace{-4.5pt}
\centerline{\includegraphics[width=\textwidth]{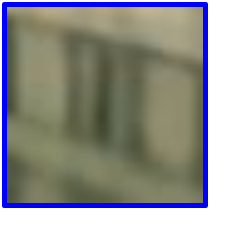}}
\vspace{-8pt}
\centerline{\hspace{-4pt}UDVD}
\vspace{-2pt}
\centerline{\hspace{-4pt}(Ours)}
\end{minipage}
\hspace{-7.7pt}
\begin{minipage}{0.062\textwidth}
\vspace{-2pt}
\centerline{\includegraphics[width=\textwidth]{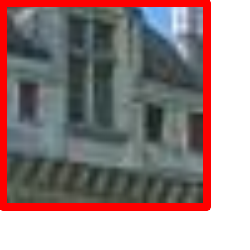}}
\vspace{-4.5pt}
\centerline{\includegraphics[width=\textwidth]{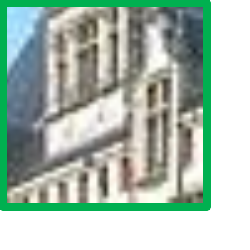}}
\vspace{-4.5pt}
\centerline{\includegraphics[width=\textwidth]{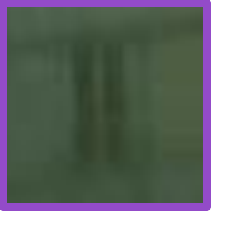}}
\vspace{-4.5pt}
\centerline{\includegraphics[width=\textwidth]{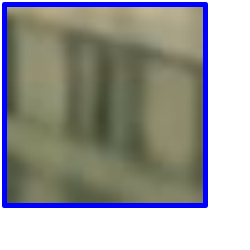}}
\vspace{-8pt}
\centerline{\hspace{-4pt}HR}
\end{minipage}
\caption{The SISR problem with variational degradations. Image “castle (102061)” in BSD100~\cite{BSD100} with scale factor 2 is used. The HR image applies Gaussian blur with variant kernel width $\varepsilon$, bicubicly downsampling and white Gaussian noise with variant level $\sigma$ to generate LR image.}
\vspace{-15pt}
\label{introimg}
\end{figure}

Several CNN base methods~\cite{ZSSR,SRMD,SFTMD,CCGAN,InternalGAN} are proposed in the context of SISR with multiple degradations. These methods address this problem with very diverse settings and formulations. Shocher \etal~\cite{ZSSR} train a small image-specific network to deal with different degradations for a certain image. In this approach, an individual network has to be trained whenever there is variation of degradations, \eg, different degrading effects across images. To address the problem of variational degradations, Zhang \etal~\cite{SRMD} propose SRMD and train a single network to handle multiple variations, including blur and noise. Note that the types of degrading effect are predefined, which is also known as non-blind setting. To the best of our knowledge, SFTMD~\cite{SFTMD} is one of the most recent works for blind setting, which proposes Spatial Feature Transform (SFT) and Iterative Kernel Correction (IKC) to deal with a limited set of blind degradations. Following the most related work, SRMD~\cite{SRMD}, this paper adopts the same setting and formulation in which a single unified network is trained for variational degradations.

To handle variations of degrading effects, a unified network is expected to accommodate two types of variations, cross-image variations (inter-image) and spatial variations (intra-image). Fig.~\ref{introimg} illustrates an example of the problem of variational degradations in SISR. In Fig.~\ref{introimg}, different degrading effects are applied to different regions of a benchmarking image in BSD100~\cite{BSD100}. Compare to the methods trained with a fixed degradation setting, RCAN~\cite{RCAN} and ZSSR~\cite{ZSSR}, the proposed method achieves similar quality (\textcolor{red}{red} patches) while effectively adapt to other variations (\textcolor{green}{green}, \textcolor{purple}{purple} and \textcolor{blue}{blue} patches in Fig.~\ref{introimg}). On the other hand, in RCAN~\cite{RCAN} and ZSSR~\cite{ZSSR}, unsatisfying quality can be observed due to its unawareness of variations. Further discussions and comparisons with a wide range of existing works will be addressed in Section~\ref{Experiments}.

Different from the most directly comparable approach, SRMD~\cite{SRMD}, this paper exploits dynamic convolutions to better solve the non-blind SISR problem with variational degradations. Dynamic convolution is a far more flexible operation than the standard one. A standard convolution learns kernels that minimize the error across all pixel locations at once. While, dynamic convolution uses per-pixel kernels generated by the parameter-generating network~\cite{DFN}. Moreover, the kernels of standard convolution are content-agnostic which are fixed after training. In contrast, the dynamic ones are content-adaptive which adapt to different input even after training. By the aforementioned properties, dynamic convolution demonstrates itself a better alternative to handle variational degradations. In this paper, we incorporate dynamic convolutions and propose a Unified Dynamic Convolutional Network for Variational Degradations (UDVD).

The contributions of this work are summarized as follows: (1) We propose UDVD, a unified dynamic convolutional network for non-blind SISR with variational degradations. (2) We further propose two types of dynamic convolutions to improve performance. And we integrate multistage loss to gradually refine images throughout the consecutive dynamic convolutions. (3) We perform comprehensive analysis of the performance impact of dynamic convolutions and investigate a number of configurations of dynamic convolutions. Extensive experiments show that the proposed UDVD achieves favorable or comparable performance on both synthetic and real images.

This paper is organized as follows. Section~\ref{Related Work} discusses related works. Section~\ref{Method} introduces the proposed UDVD and its implementation details. Section~\ref{Experiments} presents experimental results. Conclusions are discussed in Section~\ref{Conclusion}.

\section{Related Work}
\label{Related Work}
\paragraph{Single Image Super-Resolution.} In the past few years, various CNNs based techniques had been proposed for SISR. Among these works, SRCNN~\cite{SRCNN} was the first to adopt a three-layer CNN architecture and achieved superior performance against the previous non-CNN works. Inspired by SRCNN, Kim \etal. proposed a deeper network VDSR~\cite{VDSR} which contains 20 convolution layers and added global residual connection for residual learning~\cite{ResNet}. The authors in~\cite{DRCN,DRRN,MemNet} investigated the use of recursive blocks to increase the depth with parameter sharing. However, the bicubicly interpolated LR images used in these methods cause additional computation cost. To address this problem, FSRCNN~\cite{FSRCNN} and ESPCNN~\cite{ESPCNN} directly mapped LR images to HR images by adding transpose convolution layers and sub-pixel convolution layers at the end of the network. To cope with the ever decreasing of input resolution, EDSR~\cite{EDSR} and RDN~\cite{RDN} leveraged an even larger model. DBPN~\cite{DBPN} further proposed iterative upsample and downsample unit for large scaling factors. Recently, RCAN~\cite{RCAN} integrated residual-in-residual structure and channel attention mechanism to weight channel-wise features.

\vspace{-10pt}
\paragraph{Multiple Degradations.} Among the studies of SISR, most of the works were trained on a single and fixed degradation \eg, bicubic downsampling, which strongly limits the applicability in practical scenarios. In the context of multiple and diverse degradations, Shocher \etal~\cite{ZSSR} trained a small image-specific network to deal with different degradations for a certain image. The authors in~\cite{CCGAN,InternalGAN} used Generative Adversarial Networks (GANs) to tackle degradations in an unsupervised way. Zhang \etal~\cite{SRMD} proposed SRMD, a single network to handle multiple degradations, including  blur and noise. Gu \etal~\cite{SFTMD} proposed SFTMD, and applied Spatial Feature Transform (SFT) and Iterative Kernel Correction (IKC) for a subset of blind degradations. Different from the most directly comparable approach, SRMD~\cite{SRMD}, we exploit dynamic convolutions to better solve the SISR problem with variational degradations. When compared to SRMD~\cite{SRMD}, the proposed method achieves favorable performance on both synthetic and real images.

\vspace{-10pt}
\paragraph{Dynamic Kernel Network.} In recent researches, dynamic kernels in convolution layer had been widely-used in many applications. Brabandere \etal~\cite{DFN} firstly exploited parameter-generating network to generate dynamic kernels for every pixel. Dynamic kernel network had made the trained network more flexible and gained successes in various applications, such as denoising~\cite{KPN1, KPN3, KPN2, DKPN}, video super-resolution~\cite{DUF} and video interpolation~\cite{VI1, VI2}. This paper adopts a simple fully-convolution backbone to predict per-pixel kernels for dynamic convolutions. With the experiments in Section~\ref{Experiments}, this paper validates that the flexibility of dynamic convolutions can be used to tackle variational degradations of SISR.

\begin{figure*}[t]
\centering
\includegraphics[height=0.334\textheight]{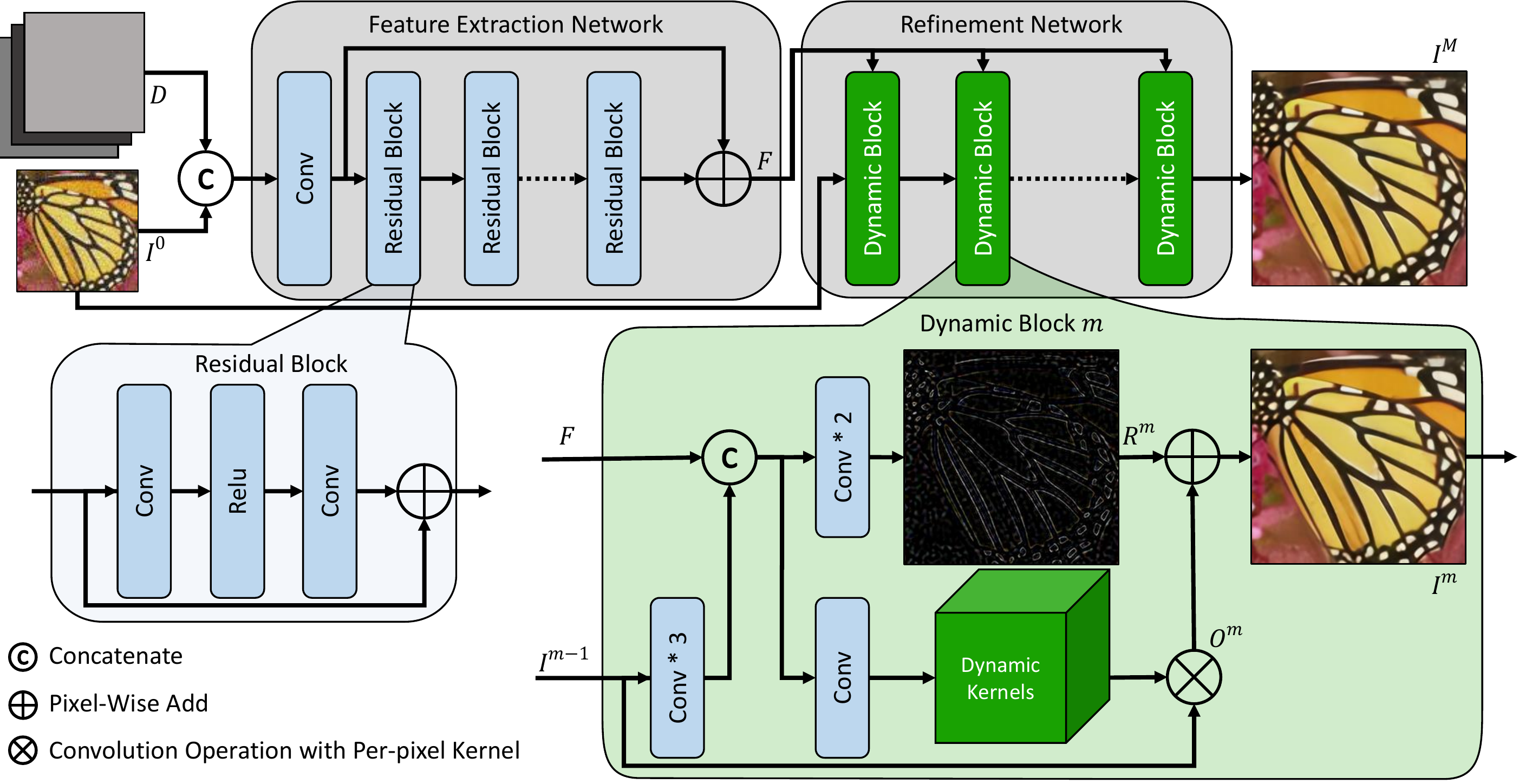}
\caption{The network architecture of the proposed UDVD framework.}
\label{framework}
\vspace{-15pt}
\end{figure*}

\section{Proposed Method}
\label{Method}
In this section, the problem of variational degradations of SISR is formulated. We then introduce the architecture and implementation details of the proposed UDVD. Next, the design of different types of dynamic convolutions and multistage loss are further discussed.

\subsection{Problem Formulation}
This paper focuses on SISR with the degrading effects, including blurring, noise and downsampling. These degrading effects can simultaneously happen to a practical use case~\cite{SISR}. The degradation process is formulated as:
\begin{equation}
I_{LR} = (I_{HR} \otimes k)\downarrow_{s} + \ n ,
\label{degrade_equation}
\end{equation}
where $I_{HR}$ and $I_{LR}$ indicates HR and LR image respectively, $k$ represents a blur kernel, $n$ stands for additive noise, $\otimes$ marks convolution operation, and $\downarrow_{s}$ denotes downsampling operation with scale factor $s$. We consider Isotropic Gaussian blur kernel, which is one of the widely used kernels in recent studies~\cite{SRMD,SFTMD,SISR}. For additive noise, most of the studies adopt Additive White Gaussian Noise (AWGN) with covariance (noise level)~\cite{SRMD,SFTMD}. Bicubic downsampler is considered for downsampling operation. By controlling the parameters of degrading effects, one can synthesize more realistic degradations for SISR training.

\vspace{-10pt}
\paragraph{Non-blind setting.}
\label{non-blind}
In this paper, a non-blind setting is adopted.  Assume given ground truth degradations, non-blind results provide the upper bounds for blind methods in which the degradations are estimated.  Such bounding observation are supported as shown by Table 2 of~\cite{ZSSR}, Table 1 of~\cite{SFTMD} and Table 1 of \cite{InternalGAN}.  As mentioned above, improvements in non-blind setting elevate the performance upper bound for blind methods \cite{SFTMD, InternalGAN}. Any degradation estimation methods can be prepended to extend our method on blind setting.

\subsection{Unified Dynamic Convolutional Network for Variational Degradation (UDVD)}
The framework of the UDVD is illustrated in Fig.~\ref{framework}. The framework consists of a feature extraction network and a refinement network. The feature extraction network manages to extract high-level features of the input image, such as global context, local details and so on. The refinement network is then learn to enhance and upsample the image together with the extracted high-level features.

\vspace{-10pt}
\paragraph{Training with Variational Degradations.} 
Given a HR image, the degrading process is executed as follows, applying isotropic Gaussian blur kernel of size $p\times p$, bicubicly downsampling the image, and finally adding AWGN with noise level $\sigma$. The generated LR image is of size $C\times H\times W$, where $C$ denotes the number of channels, $H$ and $W$ denote the height and width of the image. Similar to~\cite{SRMD}, we project the blur kernel to a $t$-dimensional vector by using Principal Component Analysis (PCA) technique. We then concatenate an extra dimension of noise level $\sigma$ to get a $(1+t)$ vector. Such vector is then stretched to get a degradation map $D$ of size $(1+t)\times H\times W$. Last, we concatenate the LR image $I^{0}$ with the degradation map $D$ of size $(C+1+t)\times H\times W$ as input for UDVD. Note that $t$ is set to $15$ by default.

\vspace{-10pt}
\paragraph{Feature Extraction Network.} In UDVD, the input is first forwarded to the feature extraction network to extract high-level features. And then, the high-level features and the input image are sent to the refinement network to generate HR image. The feature extraction network contains $N$ residual blocks which are composed of convolutions and Rectified Linear Units (ReLU), as shown in Fig.~\ref{framework}. In this work, the kernel size of convolution layers is set to $3 \times 3$, and the channels is set to 128.  

\begin{figure}[t]
\centering
\footnotesize
\begin{minipage}{0.45\textwidth}
\centerline{\includegraphics[width=0.8\linewidth]{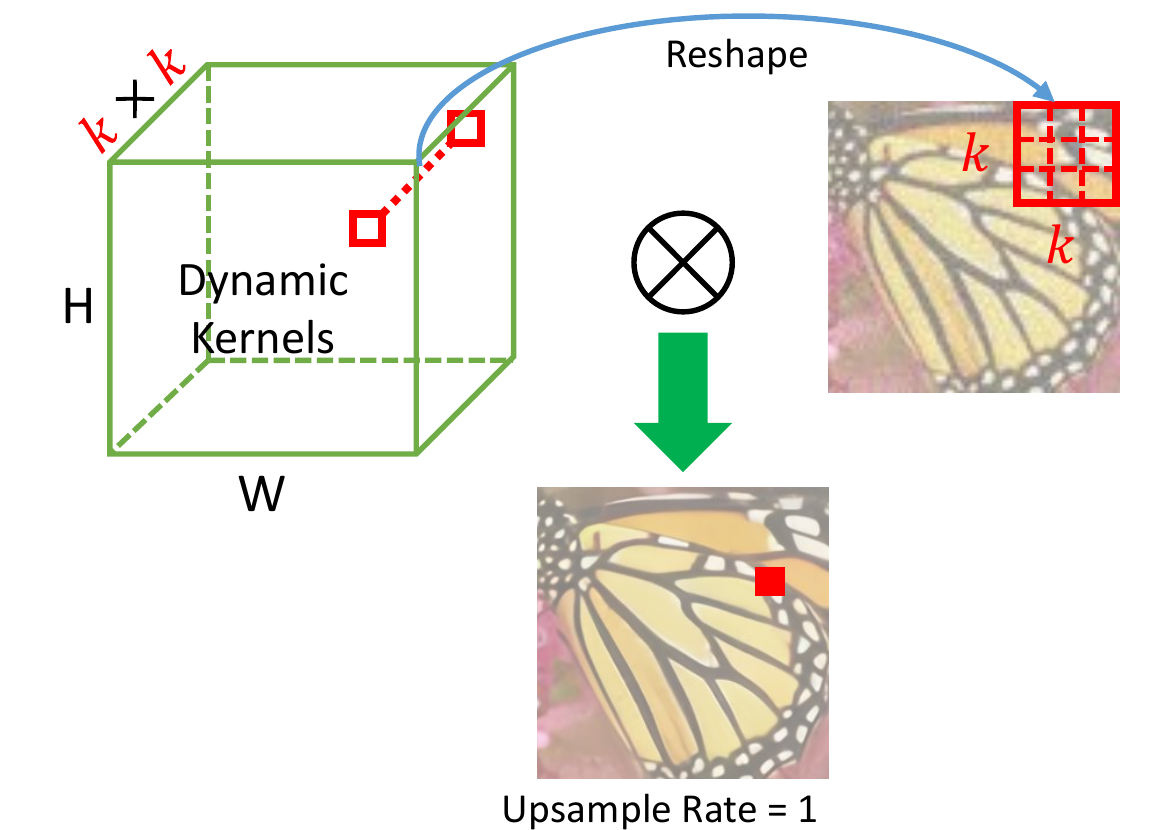}}
\vspace{-2pt}
\centerline{(a) Typical Dynamic Convolution}
\vspace{2pt}
\centerline{\includegraphics[width=0.8\linewidth]{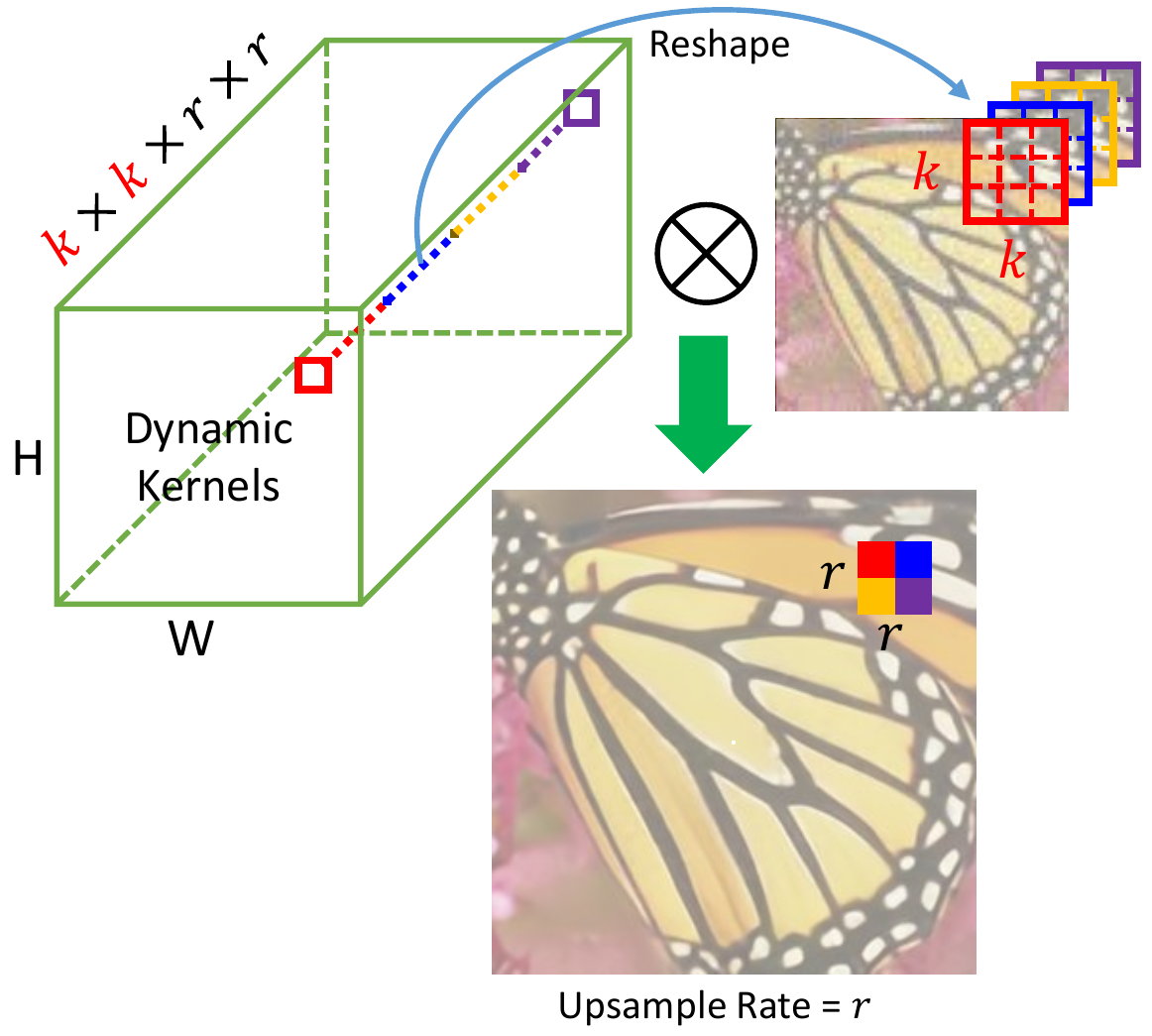}}
\vspace{-2pt}
\centerline{(b) Dynamic Convolution with Upsampling}
\vspace{5pt}
\end{minipage}
\caption{Types of dynamic convolutions. (a) A typical dynamic convolution refines the image quality while keeps resolution. (b) A dynamic convolution further upsamples the image with a specific upsample rate $r$.}
\label{dconv}
\vspace{-15pt}
\end{figure}

\vspace{-10pt}
\paragraph{Refinement Network.} 
With the extracted feature maps, the refinement network further allocates $M$ dynamic blocks for feature transformation. Note that a dynamic block can be optionally extended to perform upsampling with a specific rate $r$. The implementation details of dynamic convolutions will be covered in Section~\ref{Types of Dynamic Convolution}. In a dynamic block $m$, as illustrated in Fig.~\ref{framework}, the input image $I^{m-1}$ is sent to three $3\times 3$ convolution layers with 16, 16 and 32 channels respectively, and then concatenated with the high-level feature maps $F$ from the feature extraction network. The resultant feature maps are then forwarded to two paths. 

The first path is a $3\times 3$ convolution layer to predict per-pixel kernels. The generated per-pixel kernels are stored in a tensor with $k \times k$ in channel dimension, where $k$ is kernel size for per-pixel kernels (Fig.~\ref{dconv}(a)). When upsampling is of interested, the channel dimension is $k\times k\times r\times r$, where $r$ is upsample rate (Fig.~\ref{dconv}(b)). Note that $k$ is set to $5$ for the default setting. The predicted per-pixel kernels are then used to perform dynamic convolution operation on $I^{m-1}$ to generate the output $O^{m}$. 

The second path contains two $3 \times 3$ convolution layer with 16 and 3 channels to generate the residual image $R^{m}$ for enhancing high frequency details as describe in~\cite{DUF}. The residual image $R^{m}$ is then added to the output of dynamic convolution operation $O^{m}$ for output image $I^{m}$.  Note that sub-pixel convolution layer is used to align the resolutions between paths.

\subsection{Types of Dynamic Convolutions}
\label{Types of Dynamic Convolution}
Fig.~\ref{dconv} illustrates two types of dynamic convolutions. In general, typical dynamic convolutions are used when input and output resolution are identical, shown in Fig.~\ref{dconv}(a). Depending on use cases, upsampling can also be integrated into dynamic convolution as shown in Fig.~\ref{dconv}(b).

\vspace{-5pt}
\paragraph{Dynamic Convolution.} In a typical dynamic convolution, convolutions are conducted by using per-pixel kernels $K$ of kernel size $k \times k$. Such operation can be expressed as:
\vspace{-5pt}
\begin{equation}
I_{out}(i, j) = \sum_{u=-\Delta}^{\Delta} \sum_{v=-\Delta}^{\Delta} K_{i, j}(u, v) \cdot I_{in}(i-u, j-v) ,
\label{d_equation}
\vspace{-2pt}
\end{equation}
where $I_{in}$ and $I_{out}$ represent input and output image respectively. $i$ and $j$ are the coordinates in image, $u$ and $v$ are the coordinates in each $K_{i,j}$. Note that $\Delta = \lfloor k/2 \rfloor$. These per-pixel kernels perform weighted sum across nearby pixels and enhance the image quality pixel by pixel. In default setting, there are $H\times W$ kernels and the corresponding weights are shared across channels. By introducing an additional dimension $C$ with Eq.~\ref{d_equation}, dynamic convolution can be extended for independent weights across channels.

\vspace{-5pt}
\paragraph{Integration with Upsampling.} To integrate with upsampling, $r^{2}$ convolutions are conducted on the same corresponding patch to create $r\times r$ new pixels. The mathematical form of such operation is defined as:
\vspace{-5pt}
\begin{equation}
\begin{aligned}
I_{out}&(i\times r+x, j\times r+y) =\\
&\sum_{x=0}^{r} \sum_{y=0}^{r} \sum_{u=-\Delta}^{\Delta} \sum_{v=-\Delta}^{\Delta} K_{i,j,x,y}(u, v) \cdot I_{in}(i-u, j-v) ,
\label{u_equation}
\end{aligned}
\end{equation}
where $x$ and $y$ are in the coordination of each $r \times r$ output block $(0 \leq x, y \leq r - 1)$. Here, the resolution of $I_{out}$ is $r$ times the resolution of $I_{in}$. We exploit $r^{2}HW$ kernels to generate $rH\times rW$ pixels as $I_{out}$. When integrated with upsampling, the weights are shared across channels to avoid the curse of dimensionality~\cite{DCN}.

\subsection{Multistage Loss}
\label{Multistage Loss}
Similar to previous work~\cite{MSLapSRN, PGAN}, we adopt a multistage loss at the outputs of dynamic blocks. The losses are calculated in between the HR image $I_{HR}$ and the intermediate images at each dynamic block. The loss is defined as:
\begin{equation}
Loss = \sum_{m=1}^{M} F(I^{m}, I_{HR})
\label{equation4}
\end{equation}
where $M$ is the number of dynamic blocks and $F$ is loss function such as $L_{2}$ loss and perceptual loss. To obtain a high quality resultant image, we then minimize the sum of the losses from each dynamic block.

\section{Experiments}
\label{Experiments}
In this section, we discuss the experimental results and setups. Section~\ref{Datasets and Training Setups} elaborates the details of dataset and training setups. Different configurations of the proposed UDVD are compared in Section~\ref{Comparison of UDVD Configurations}.  Section~\ref{Visualizing Dynamic Kernels} illustrates the learned dynamic kernels. Section~\ref{Experiments on Synthetic Images} evaluates UDVD with different applicational settings using synthetic images. Real image evaluations are in Section~\ref{Experiments on Real Images}. 

\subsection{Datasets and Training Setups}
\label{Datasets and Training Setups}
We collect high-quality 2K images from DIV2K~\cite{DIV2K} and Flickr2K~\cite{Flickr2K} for training. The degraded images are synthesized according to Eq.~\ref{degrade_equation}. As listed in Eq.~\ref{degrade_equation}, the process applies a sequence of degrading effects on high-quality images, saying blurring, bicubicly downsampling and then adding noise. Following previous works~\cite{SRMD, SFTMD}, we use isotropic Gaussian blur kernels. The range of kernel width is set to $[0.2,\ 3.0]$, and the kernel size is fixed to $15\times15$. For noise, we use AWGN with noise level in range $[0,\ 75]$. Uniform sampling is used to generate all the parameters.

During training, degraded LR images are cropped into patches of size $48\times48$. While the corresponding HR images are cropped into patches of $96\times96$, $144\times144$ and $192\times192$ for scale factors 2, 3, and 4, respectively. The patches are augmented by randomly horizontal flipping, vertical flipping, and $90^{o}$ rotating. We set the mini-batch size to 32. Adam optimizer~\cite{ADAM} is used together with the multistage L2 loss introduced in Section~\ref{Multistage Loss}. The learning rate is initialized to $10^{-4}$ and decreased by half for every $2\times10^5$ steps. We conduct all the experiments on a server equipped with NVIDIA RTX 2080 Ti GPU. 

We validate UDVD on datasets Set5~\cite{Set5}, Set14~\cite{Set14}, BSD100~\cite{BSD100} and real images~\cite{chip, frog}. Same as in~\cite{RDN, RCAN, SRMD}, all models are trained on RGB space. The evaluations of PSNR and SSIM metrics are on the Y channel in YCbCr.

\begin{table}[t]
\small
\centering
\begin{tabular}{lcc}
Methods                 & PSNR          & SSIM          \\ \hline\hline
Baseline                 & 29.45         & 0.8350         \\
UDVD\_\textbf{U}            & 29.56         & 0.8384         \\
UDVD\_\textbf{DU}           & 29.58         & 0.8393         \\
UDVD\_\textbf{UD}           & 29.61         & 0.8400         \\
UDVD\_\textbf{UDD} w/o multistage loss & 29.58         & 0.8385         \\
UDVD\_\textbf{UDD}           & \textcolor{red}{29.67} & \textcolor{red}{0.8410} \\ \hline
\end{tabular}
\vspace{2pt}
\caption{Average PSNR and SSIM values for various UDVD configurations on Set5. Degradation parameters include scaling factor $\times3$, kernel width $1.3$ and noise level $15$ . The best results are highlighted in \textcolor{red}{red} color.}
\label{configurations}
\vspace{-15pt}
\end{table}

\subsection{Comparison of UDVD Configurations}
\label{Comparison of UDVD Configurations}
Table~\ref{configurations} summarizes the quantitative comparisons of different UDVD configurations. The baseline configuration contains only the feature extraction network (15 residual blocks) and a sub-pixel layer. Whereas, in the refinement network, different configurations of dynamic convolutions are compared. In these configurations, \textbf{D} represents the block containing a typical dynamic convolution (as shown in Fig.~\ref{dconv}(a)), and \textbf{U} marks the integration of upsampling (listed in Fig.~\ref{dconv}(b)). As in Table~\ref{configurations}, dynamic blocks improve the performance over the baseline configuration. Especially, the \textbf{UD} configuration achieves better performance over its counterparts. The ablation experiments in Table~\ref{configurations} also demonstrate the effectiveness of the proposed multistage loss. In the following of this paper, we use UDVD\_\textbf{UDD} for scale factors 2 and 3. While, UDVD\_\textbf{UUDD} (two separated upsampling steps) is used for scale factor 4.

\begin{figure}[t]
\centering
\footnotesize
\begin{minipage}{0.18\textwidth}
\boldmath
\centerline{\includegraphics[width=\textwidth]{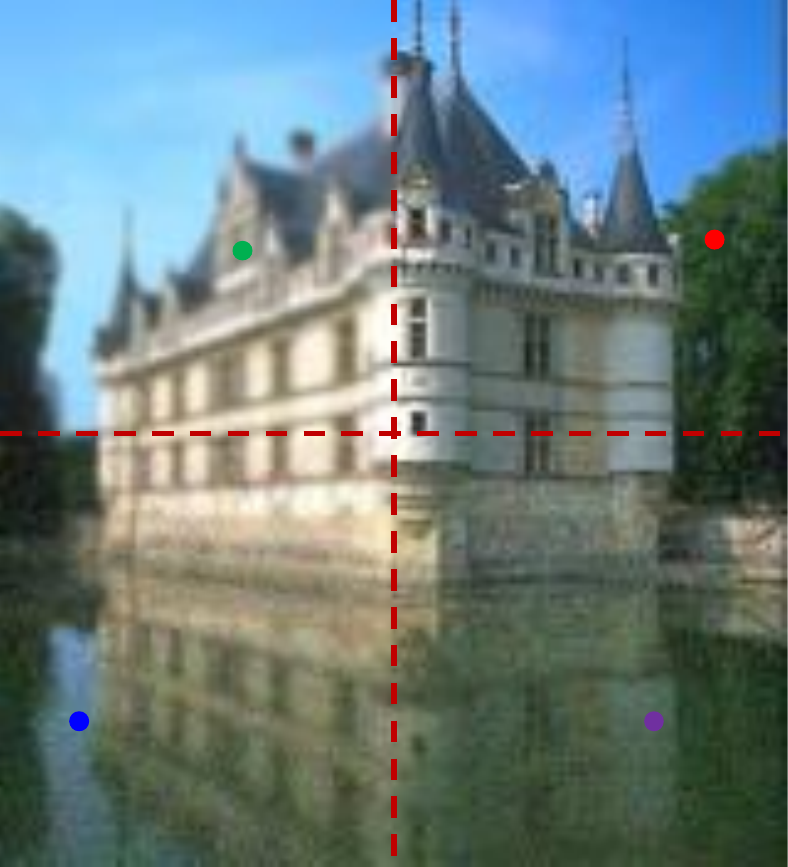}}
\vspace{-100pt}
\centerline{\textcolor{white}{\textbf{$\varepsilon$=1.6, $\sigma$=0 \hspace{8pt} $\varepsilon$=0.2,$\sigma$=0}}}
\vspace{80pt}
\centerline{\textcolor{white}{\textbf{$\varepsilon$=1.6,$\sigma$=10 \hspace{5pt} $\varepsilon$=0.2,$\sigma$=10}}}
\centerline{Degraded Image}
\end{minipage}
\begin{minipage}{0.25\textwidth}
\centerline{\includegraphics[width=\textwidth]{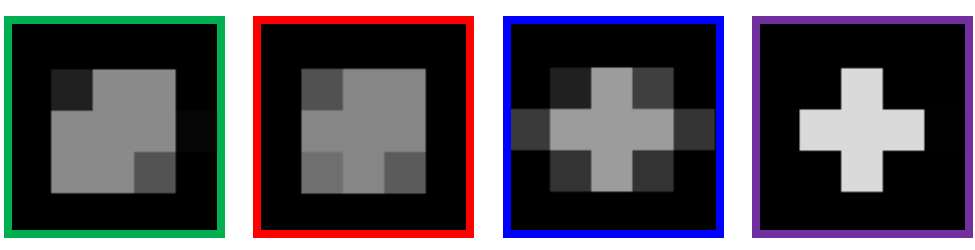}}
\vspace{-3pt}
\centerline{(a) Dynamic kernels of original image}
\centerline{\includegraphics[width=\textwidth]{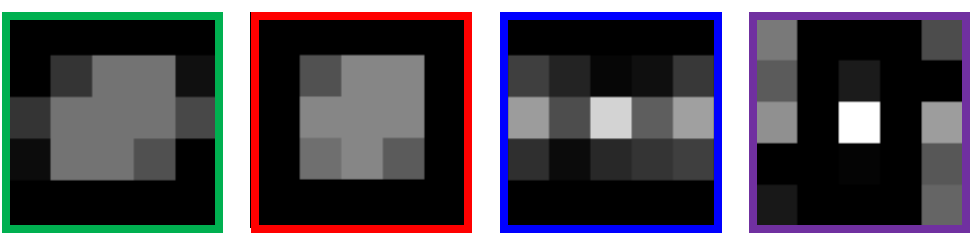}}
\vspace{-3pt}
\centerline{(b) Dynamic kernels of degraded image}
\centerline{\includegraphics[width=\textwidth]{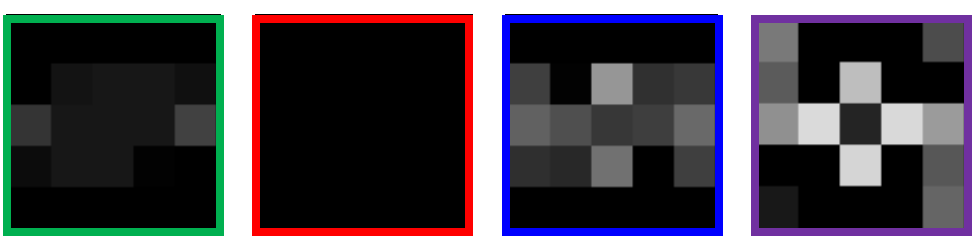}}
\vspace{-3pt}
\centerline{(c) Absolute difference between (a), (b)}
\vspace{2pt}
\end{minipage}
\begin{minipage}{0.039\textwidth}
\centerline{\includegraphics[width=\textwidth]{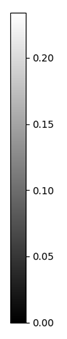}}
\end{minipage}
\caption{The predicted per-pixel kernels in second dynamic block of UDVD. (a) The kernels learned from lr image. (b) The kernels learned from image with spatially variant degradation of Gaussian blur kernel width $\varepsilon$ and noise level $\sigma$. (c) The absolute difference between (a) and (b).}
\vspace{-15pt}
\label{dynamic_kernel}
\end{figure}

\subsection{Visualizing Dynamic Kernels}
\label{Visualizing Dynamic Kernels}
UDVD generates dynamic kernels to adapt to both image contents and different degradations. Fig.~{\ref{dynamic_kernel}(a)} demonstrates that different kernels are generated for different contents. Likewise, Fig.~{\ref{dynamic_kernel}(b)} shows that the generated kernels further adapt to the applied degradations. As shown in Fig.~{\ref{dynamic_kernel}(c)}, the adaptation behavior is varying among different degradations regardless of contents. These observations confirm that UDVD is capable of handling spatial variations by generating dynamic kernels considering the spatial differences of content and degradation.

\subsection{Experiments on Synthetic Images}
\label{Experiments on Synthetic Images}
We evaluate the proposed UDVD with a number of applicational settings, covering both variational and fixed degradations. Synthetic images are generated for these settings. Note that we compared UDVD with a comprehensive set of other methods in non-blind setting only. Details are elaborated in the following paragraphs.

\begin{table*}[t]
\small
\centering
\begin{tabular}{lccccccccccc}
\hline
\multirow{2}{*}{Methods} & \multirow{2}{*}{Kernel width} & \multirow{2}{*}{Noise level} & \multicolumn{3}{c}{Set5}                         & \multicolumn{3}{c}{Set14}                        & \multicolumn{3}{c}{BSD100}                        \\
             &                &               & $\times$2       & $\times$3       & $\times$4       & $\times$2       & $\times$3       & $\times$4       & $\times$2       & $\times$3       & $\times$4       \\ \hline\hline
RDN~\cite{RDN}   & \multirow{5}{*}{0.2}     & \multirow{5}{*}{15}     & 26.23         & 25.57         & 24.48         & 25.44         & 24.40         & 23.45         & 25.03         & 24.04         & 23.13         \\
RCAN~\cite{RCAN}  &                &               & 26.05         & 25.46         & 24.83         & 25.3          & 24.29         & 23.64         & 24.95         & 23.92         & 23.33         \\
IRCNN~\cite{IRCNN}    &                &               & 32.60         & 30.08         & 28.35         & -           & -           & -           & -           & -           & -           \\
SRMD~\cite{SRMD}     &                &               & 32.76         & 30.43         & 28.79         & 30.14         & 27.82         & 26.48         & 29.23         & 27.11         & 25.95         \\
UDVD          &                &               & \textcolor{red}{32.96} & \textcolor{red}{30.68} & \textcolor{red}{29.04} & \textcolor{red}{30.43} & \textcolor{red}{28.14} & \textcolor{red}{26.82} & \textcolor{red}{29.38} & \textcolor{red}{27.27} & \textcolor{red}{26.08} \\ \hline
RDN~\cite{RDN}   & \multirow{5}{*}{1.3}     & \multirow{5}{*}{15}     & 25.01         & 24.98         & 24.33         & 24.08         & 23.92         & 23.39         & 23.85         & 23.67         & 23.09         \\
RCAN~\cite{RCAN}  &                &               & 24.9          & 24.94         & 24.58         & 24.04         & 23.88         & 23.53         & 23.84         & 23.62         & 23.26         \\
IRCNN~\cite{IRCNN}    &                &               & 29.96         & 28.68         & 27.71         & -           & -           & -           & -           & -           & -           \\
SRMD~\cite{SRMD}     &                &               & 30.98         & 29.43         & 28.21         & 28.34         & 27.05         & 26.06         & 27.52         & 26.45         & 25.63         \\
UDVD          &                &               & \textcolor{red}{31.16} & \textcolor{red}{29.67} & \textcolor{red}{28.43} & \textcolor{red}{28.63} & \textcolor{red}{27.36} & \textcolor{red}{26.37} & \textcolor{red}{27.64} & \textcolor{red}{26.58} & \textcolor{red}{25.74} \\ \hline
RDN~\cite{RDN}   & \multirow{5}{*}{2.6}     & \multirow{5}{*}{15}     & 23.18         & 23.28         & 23.07         & 22.34         & 22.40         & 22.31         & 22.44         & 22.52         & 22.35         \\
RCAN~\cite{RCAN}  &                &               & 23.13         & 23.29         & 23.24         & 22.34         & 22.41         & 22.42         & 22.47         & 22.5          & 22.48         \\
IRCNN~\cite{IRCNN}    &                &               & 26.44         & 25.67         & 24.36         & -           & -           & -           & -           & -           & -           \\
SRMD~\cite{SRMD}     &                &               & 28.48         & 27.55         & 26.82         & 26.18         & 25.58         & 25.06         & 25.81         & 25.29         & 24.86         \\
UDVD          &                &               & \textcolor{red}{28.73} & \textcolor{red}{27.80} & \textcolor{red}{26.98} & \textcolor{red}{26.48} & \textcolor{red}{25.87} & \textcolor{red}{25.33} & \textcolor{red}{25.93} & \textcolor{red}{25.41} & \textcolor{red}{24.96} \\ \hline
RDN~\cite{RDN}   & \multirow{5}{*}{0.2}     & \multirow{5}{*}{50}     & 17.23         & 16.85         & 16.51         & 17.04         & 16.58         & 16.21         & 16.90         & 16.38         & 15.99         \\
RCAN~\cite{RCAN}  &                &               & 17.08         & 16.13         & 16.64         & 16.84         & 15.68         & 16.35         & 16.66         & 15.54         & 16.1          \\
IRCNN~\cite{IRCNN}    &                &               & 28.20         & 26.25         & 24.95         & -           & -           & -           & -           & -           & -           \\
SRMD~\cite{SRMD}     &                &               & 28.51         & 26.48         & 25.18         & 26.70         & 25.01         & 23.95         & 26.13         & 24.74         & 23.86         \\
UDVD          &                &               & \textcolor{red}{28.63} & \textcolor{red}{26.65} & \textcolor{red}{25.34} & \textcolor{red}{27.00} & \textcolor{red}{25.32} & \textcolor{red}{24.24} & \textcolor{red}{26.27} & \textcolor{red}{24.87} & \textcolor{red}{23.98} \\ \hline
RDN~\cite{RDN}   & \multirow{5}{*}{1.3}     & \multirow{5}{*}{50}     & 16.97         & 16.70         & 16.41         & 16.75         & 16.45         & 16.14         & 16.64         & 16.29         & 15.95         \\
RCAN~\cite{RCAN}  &                &               & 16.82         & 15.98         & 16.54         & 16.55         & 15.56         & 16.28         & 16.42         & 15.47         & 16.06         \\
IRCNN~\cite{IRCNN}    &                &               & 26.69         & 25.20         & 24.42         & -           & -           & -           & -           & -           & -           \\
SRMD~\cite{SRMD}     &                &               & 27.43         & 25.82         & 24.77         & 25.63         & 24.47         & 23.64         & 25.26         & 24.33         & 23.63         \\
UDVD          &                &               & \textcolor{red}{27.54} & \textcolor{red}{25.99} & \textcolor{red}{24.92} & \textcolor{red}{25.88} & \textcolor{red}{24.75} & \textcolor{red}{23.91} & \textcolor{red}{25.36} & \textcolor{red}{24.45} & \textcolor{red}{23.74} \\ \hline
RDN~\cite{RDN}   & \multirow{5}{*}{2.6}     & \multirow{5}{*}{50}     & 16.50         & 16.31         & 16.08         & 16.30         & 16.09         & 15.88         & 16.29         & 16.03         & 15.77         \\
RCAN~\cite{RCAN}  &                &               & 16.36         & 15.6          & 16.22         & 16.12         & 15.24         & 16.02         & 16.07         & 15.23         & 15.88         \\
IRCNN~\cite{IRCNN}    &                &               & 22.98         & 22.16         & 21.43         & -           & -           & -           & -           & -           & -           \\
SRMD~\cite{SRMD}     &                &               & 25.85         & 24.75         & 23.98         & 24.32         & 23.53         & 22.98         & 24.30         & 23.68         & 23.18         \\
UDVD          &                &               & \textcolor{red}{26.00} & \textcolor{red}{24.85} & \textcolor{red}{24.11} & \textcolor{red}{24.60} & \textcolor{red}{23.81} & \textcolor{red}{23.23} & \textcolor{red}{24.41} & \textcolor{red}{23.79} & \textcolor{red}{23.27} \\ \hline
\end{tabular}
\vspace{2pt}
\caption{Average PSNR values on variations of multiple degradations. We use the provided official code to compute the results, except IRCNN. For IRCNN, results are extracted from the publication~\cite{IRCNN}. The best results are highlighted in \textcolor{red}{red} color.} 
\label{multiple}
\vspace{-8pt}
\end{table*}

\begin{table*}[t]
\small
\centering
\begin{tabular}{lccccccccccc}
\hline
\multirow{2}{*}{Methods} & \multirow{2}{*}{Kernel width} & \multirow{2}{*}{Noise level} & \multicolumn{3}{c}{Set5}                         & \multicolumn{3}{c}{Set14}                        & \multicolumn{3}{c}{BSD100}                        \\
             &                &               & $\times$2       & $\times$3       & $\times$4       & $\times$2       & $\times$3       & $\times$4       & $\times$2       & $\times$3       & $\times$4       \\ \hline\hline
SRMD~\cite{SRMD}     & \multirow{2}{*}{[0.2, 2]}   & \multirow{2}{*}{5}      & 33.56         & 31.67         & 30.01         & 29.79         & 28.06         & 26.77         & 27.54         & 26.00         & 24.98         \\
UDVD          &                &               & \textcolor{red}{33.77} & \textcolor{red}{32.00} & \textcolor{red}{30.63} & \textcolor{red}{30.74} & \textcolor{red}{28.93} & \textcolor{red}{27.76} & \textcolor{red}{29.61} & \textcolor{red}{27.99} & \textcolor{red}{26.90} \\ \hline
SRMD~\cite{SRMD}     & \multirow{2}{*}{0.2}     & \multirow{2}{*}{[5, 50]}   & 29.70         & 27.80         & 26.53         & 28.02         & 26.20         & 24.98         & 27.39         & 25.75         & 24.81         \\
UDVD          &                &               & \textcolor{red}{30.78} & \textcolor{red}{28.61} & \textcolor{red}{27.15} & \textcolor{red}{28.90} & \textcolor{red}{26.85} & \textcolor{red}{25.62} & \textcolor{red}{27.98} & \textcolor{red}{26.20} & \textcolor{red}{25.19} \\ \hline
SRMD~\cite{SRMD}     & \multirow{2}{*}{[0.2, 2]}   & \multirow{2}{*}{[5, 50]}   & 28.53         & 27.06         & 26.05         & 26.97         & 25.62         & 24.66         & 26.46         & 25.33         & 24.59         \\
UDVD          &                &               & \textcolor{red}{29.37} & \textcolor{red}{27.74} & \textcolor{red}{26.57} & \textcolor{red}{27.63} & \textcolor{red}{26.28} & \textcolor{red}{25.23} & \textcolor{red}{26.89} & \textcolor{red}{25.71} & \textcolor{red}{24.95} \\ \hline
\end{tabular}
\vspace{3pt}
\caption{Average PSNR values on spatial variations of degradations. We use the official code of SRMD to compute its results. The best results are highlighted in \textcolor{red}{red} color.}
\label{spatially}
\vspace{-17pt}
\end{table*}

\vspace{-10pt}
\paragraph{Variations of Multiple Degradations.} The proposed UDVD are evaluated with diverse variations of degrading effects, including bicubicly downsampling, isotropic Gaussian blur kernels and AWGN. We consider scale factors 2, 3, and 4 for bicubicly downsampling. For isotropic Gaussian blur, kernel widths 0.2, 1.3 and 2.6 are considered. Noise levels 15 and 50 are considered for AWGN. Table~\ref{multiple} compares UDVD to previous works on widely-used dataset, Set5~\cite{Set5}, Set14~\cite{Set14} and BSD100~\cite{BSD100}. Due to the unawareness of multiple degradations, RDN~\cite{RDN} and RCAN~\cite{RCAN} produce unsatisfying PSNR when compared to other methods. For multiple degradations based methods, we compare UDVD to two well known methods, IRCNN~\cite{IRCNN}, and SRMD~\cite{SRMD}. As in Table~\ref{multiple}, UDVD constantly achieves better PSNR, even at the hardest test case (kernel width 2.6 and noise level 50).

\begin{figure*}[t]
\centering
\footnotesize
\begin{minipage}{0.195\textwidth}
\centerline{\includegraphics[width=\textwidth]{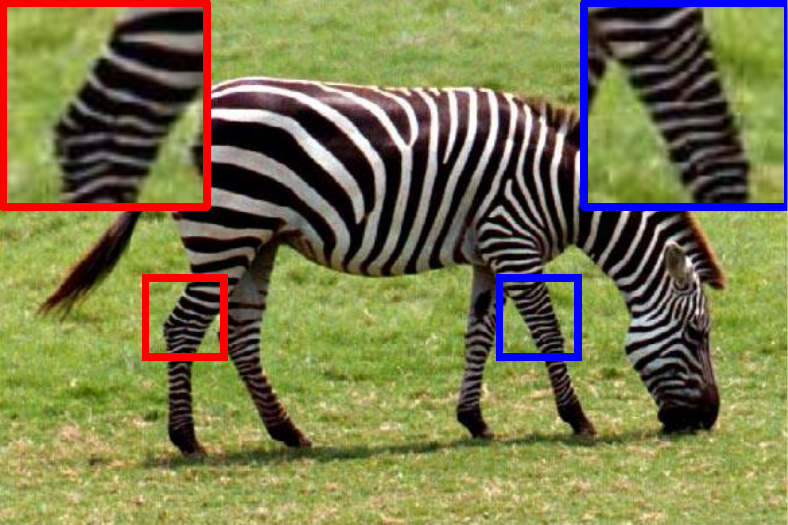}}
\vspace{-1pt}
\centerline{PSNR/SSIM}
\vspace{1pt}
\centerline{\includegraphics[width=\textwidth]{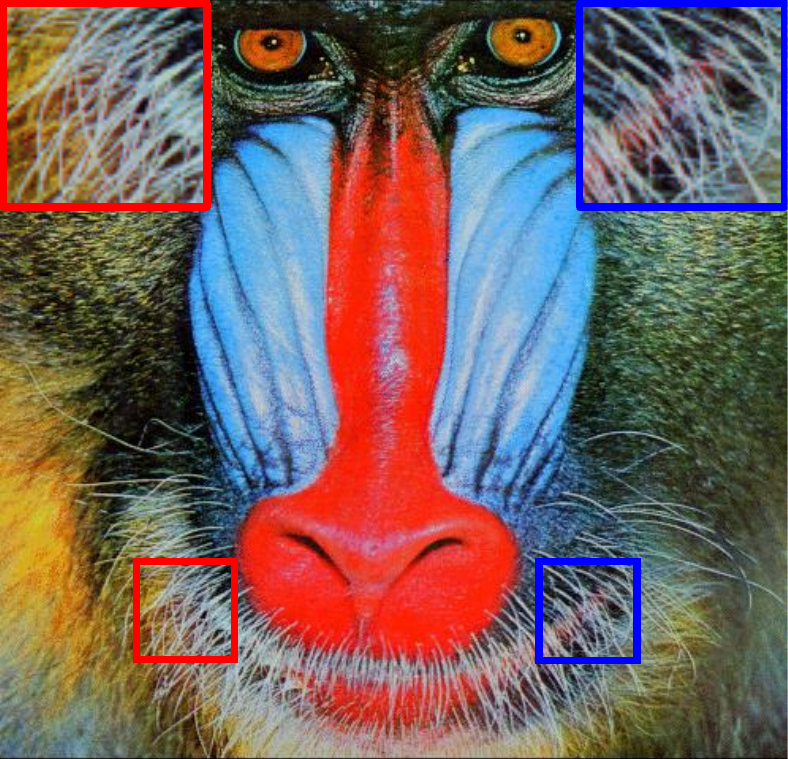}}
\vspace{-1pt}
\centerline{PSNR/SSIM}
\centerline{(a) Ground Truth}
\end{minipage}
\begin{minipage}{0.195\textwidth}
\centerline{\includegraphics[width=\textwidth]{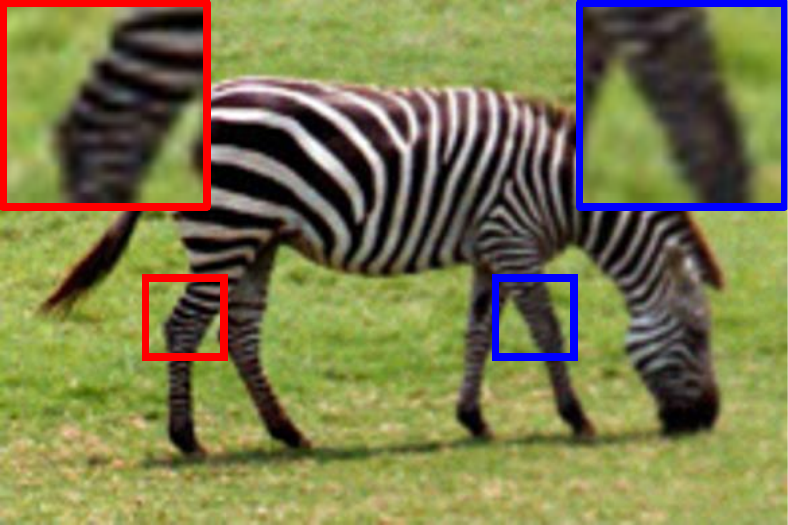}}
\vspace{-1pt}
\centerline{24.50/0.7148}
\vspace{1pt}
\centerline{\includegraphics[width=\textwidth]{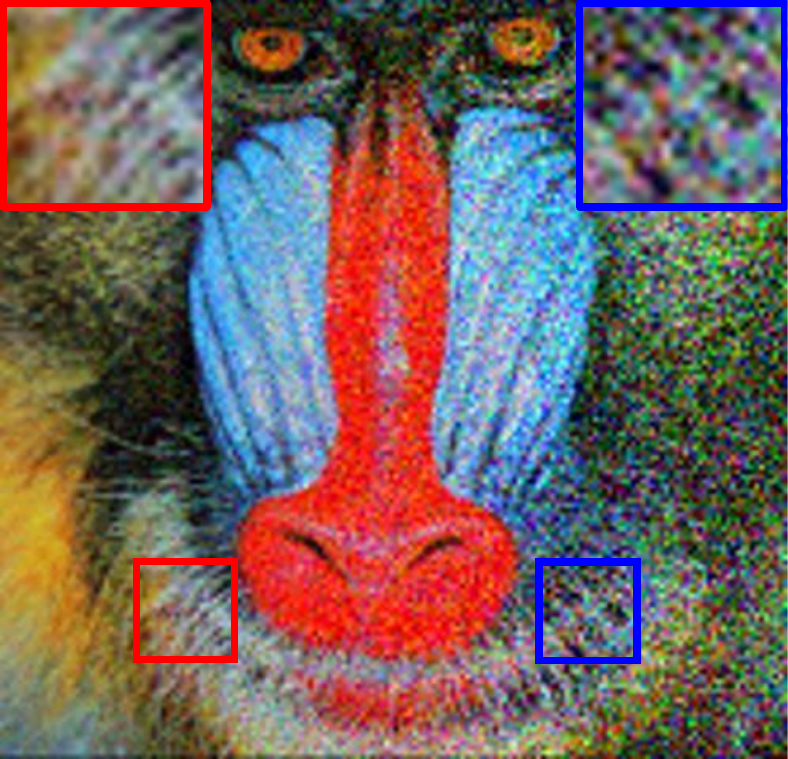}}
\vspace{-1pt}
\centerline{20.56/0.3831}
\centerline{(b) Bicubic}
\end{minipage}
\begin{minipage}{0.195\textwidth}
\centerline{\includegraphics[width=\textwidth]{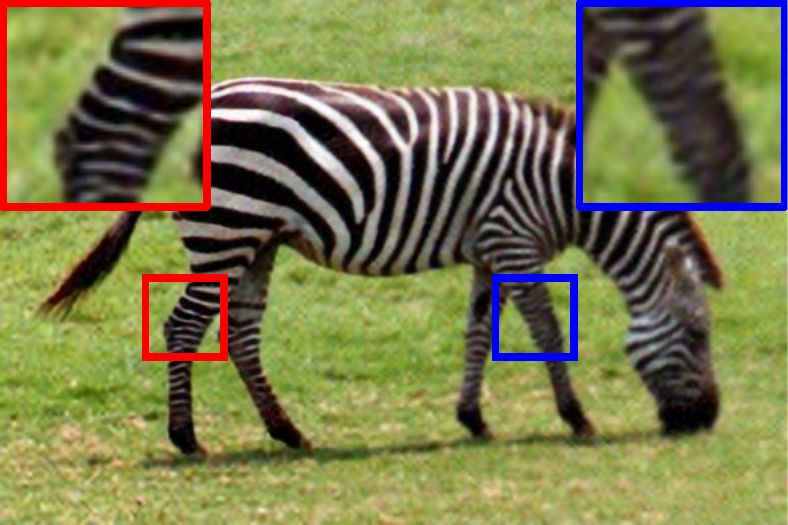}}
\vspace{-1pt}
\centerline{26.03/0.7546}
\vspace{1pt}
\centerline{\includegraphics[width=\textwidth]{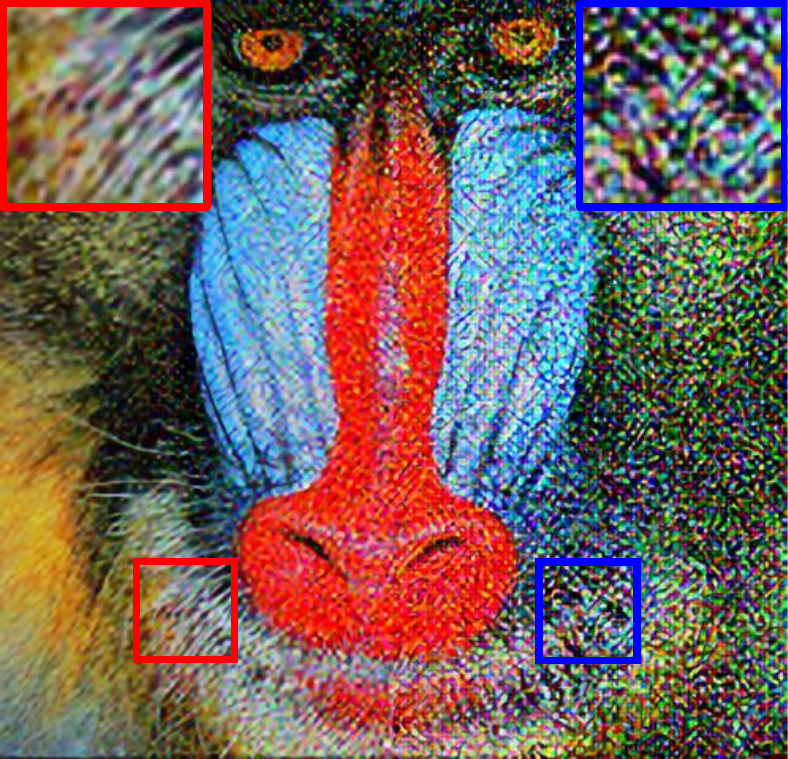}}
\vspace{-1pt}
\centerline{17.68/0.3068}
\centerline{(c) RCAN~\cite{RCAN}}
\end{minipage}
\begin{minipage}{0.195\textwidth}
\centerline{\includegraphics[width=\textwidth]{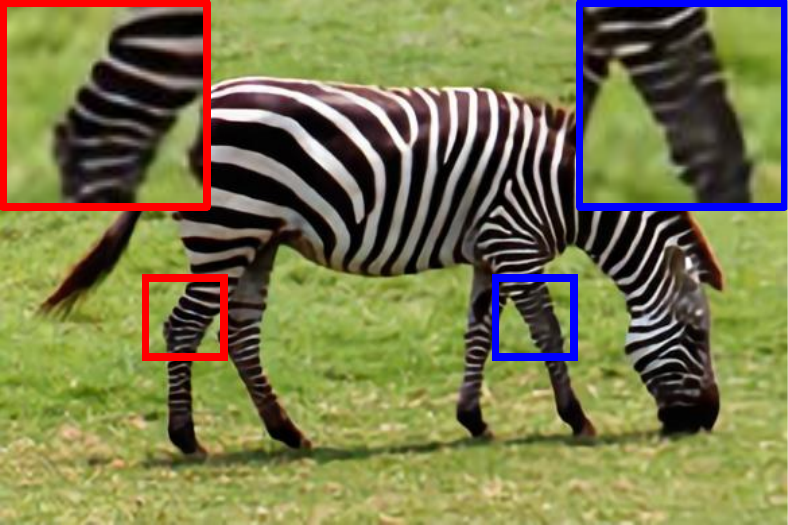}}
\vspace{-1pt}
\centerline{28.06/0.8005}
\vspace{1pt}
\centerline{\includegraphics[width=\textwidth]{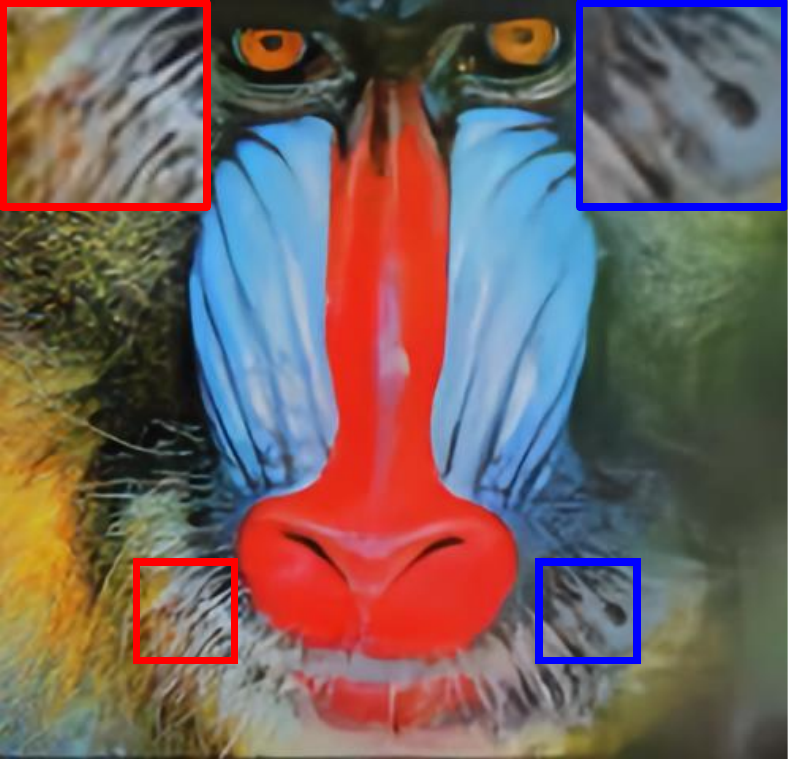}}
\vspace{-1pt}
\centerline{22.33/0.4608}
\centerline{(d) SRMD~\cite{SRMD}}
\end{minipage}
\begin{minipage}{0.195\textwidth}
\centerline{\includegraphics[width=\textwidth]{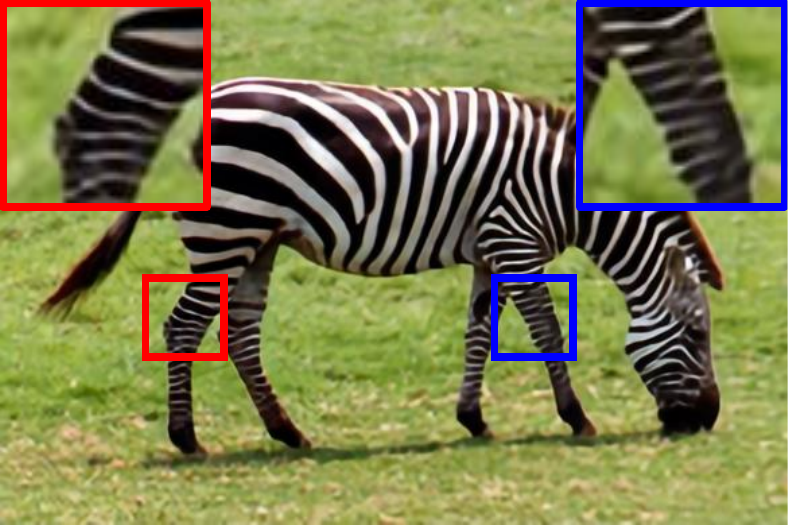}}
\vspace{-1pt}
\centerline{28.45/0.8052}
\vspace{1pt}
\centerline{\includegraphics[width=\textwidth]{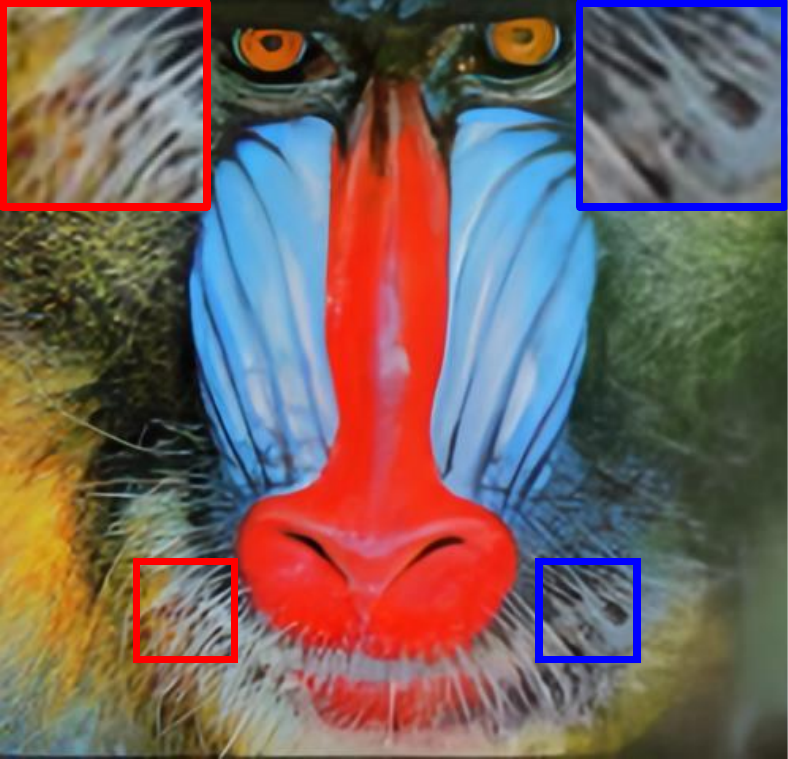}}
\vspace{-1pt}
\centerline{22.52/0.4856}
\centerline{(e) UDVD(Ours)}
\end{minipage}
\vspace{5pt}
\caption{The visual results of image \textit{“zebra”} (first row) and image \textit{“bamboo”} (second row) with scale factor 3 on spatially variant blur and noise degradations.} 
\label{SVimg}
\vspace{-8pt}
\end{figure*}

\begin{table*}[t]
\small
\centering
\begin{tabular}{lcccccccccc}
\hline
\multirow{2}{*}{Methods}  & \multirow{2}{*}{Kernel width} & \multicolumn{3}{c}{Set5}                          & \multicolumn{3}{c}{Set14}                          & \multicolumn{3}{c}{BSD100}                         \\
              &                & $\times$2        & $\times$3        & $\times$4        & $\times$2        & $\times$3        & $\times$4        & $\times$2        & $\times$3        & $\times$4        \\ \hline\hline
ZSSR~\cite{ZSSR}      & \multirow{6}{*}{0.2}     & 34.94          & 29.29          & 28.87          & 31.04          & 28.05          & 27.15          & 31.42          & 28.24          & 26.68          \\
IRCNN~\cite{IRCNN}     &                & 37.43          & 33.39          & 31.02          & -            & -            & -            & -            & -            & -            \\
SRMDNF~\cite{SRMD}     &                & 37.79          & 34.13          & 31.96          & 33.33          & 30.04          & 28.35          & 32.05          & 28.97          & 27.49          \\
SFTMD w/o SFT~\cite{SFTMD} &                & 31.74          & 30.90          & 29.40          & 27.57          & 26.40          & 26.18          & 27.24          & 26.43          & 26.34          \\
SFTMD~\cite{SFTMD}     &                & \textcolor{blue}{38.00} & \textcolor{red}{34.57} & \textcolor{red}{32.39} & \textcolor{red}{33.68} & \textcolor{red}{30.47} & \textcolor{blue}{28.77} & \textcolor{blue}{32.09} & \textcolor{blue}{29.09} & \textcolor{blue}{27.58} \\
UDVD           &                & \textcolor{red}{38.01} & \textcolor{blue}{34.49} & \textcolor{blue}{32.31} & \textcolor{blue}{33.64} & \textcolor{blue}{30.44} & \textcolor{red}{28.78} & \textcolor{red}{32.19} & \textcolor{red}{29.18} & \textcolor{red}{27.70} \\ \hline
ZSSR~\cite{ZSSR}      & \multirow{6}{*}{1.3}     & 33.37          & 28.67          & 27.44          & 31.31          & 27.34          & 26.15          & 30.31          & 27.30          & 25.95          \\
IRCNN~\cite{IRCNN}     &                & 36.01          & 33.33          & 31.01          & -            & -            & -            & -            & -            & -            \\
SRMDNF~\cite{SRMD}     &                & \textcolor{blue}{37.44} & 34.17          & 32.00          & 33.20          & 30.08          & 28.42          & \textcolor{blue}{31.98} & 29.03          & 27.53          \\
SFTMD w/o SFT~\cite{SFTMD} &                & 30.88          & 30.33          & 29.11          & 27.16          & 25.84          & 25.93          & 26.84          & 25.92          & 26.20          \\
SFTMD~\cite{SFTMD}     &                & \textcolor{red}{37.46} & \textcolor{red}{34.53} & \textcolor{red}{32.41} & \textcolor{red}{33.39} & \textcolor{red}{30.55} & \textcolor{blue}{28.82} & \textcolor{red}{32.06} & \textcolor{blue}{29.15} & \textcolor{blue}{27.64} \\
UDVD           &                & 37.36          & \textcolor{blue}{34.52} & \textcolor{blue}{32.37} & \textcolor{red}{33.39} & \textcolor{blue}{30.50} & \textcolor{red}{28.85} & 32.00          & \textcolor{red}{29.23} & \textcolor{red}{27.75} \\ \hline
ZSSR~\cite{ZSSR}      & \multirow{6}{*}{2.6}     & 29.89          & 27.80          & 27.69          & 27.72          & 26.42          & 26.06          & 27.32          & 26.47          & 25.92          \\
IRCNN~\cite{IRCNN}     &                & 32.07          & 31.09          & 30.06          & -            & -            & -            & -            & -            & -            \\
SRMDNF~\cite{SRMD}     &                & \textcolor{blue}{34.12} & 33.02          & 31.77          & \textcolor{blue}{30.25} & 29.33          & 28.26          & \textcolor{blue}{29.23} & 28.35          & 27.43          \\
SFTMD w/o SFT~\cite{SFTMD} &                & 24.22          & 28.44          & 28.64          & 22.99          & 24.19          & 25.63          & 23.07          & 24.42          & 25.99          \\
SFTMD~\cite{SFTMD}     &                & \textcolor{red}{34.27} & \textcolor{red}{33.22} & \textcolor{red}{32.05} & \textcolor{red}{30.38} & \textcolor{red}{29.63} & \textcolor{red}{28.55} & \textcolor{red}{29.35} & \textcolor{blue}{28.41} & \textcolor{blue}{27.47} \\
UDVD           &                & 33.74          & \textcolor{blue}{33.15} & \textcolor{blue}{31.99} & 30.08          & \textcolor{blue}{29.58} & \textcolor{red}{28.55} & 28.93          & \textcolor{red}{28.49} & \textcolor{red}{27.55} \\ \hline
\end{tabular}
\vspace{2pt}
\caption{Average PSNR values on noise-free degradations. We use the official code to compute the results, except SFTMD and IRCNN. For SFTMD and IRCNN, results are extracted from the publications~\cite{SFTMD, IRCNN}. The best two results are highlighted in \textcolor{red}{red} and \textcolor{blue}{blue} colors.}
\label{noisefree}
\vspace{-17pt}
\end{table*}

\vspace{-10pt}
\paragraph{Spatial Variations of Degradations.} In order to validate the advantages of dynamic convolution, we further extend the experiments to consider spatial variations of degradations. The degraded LR images are synthesized with the kernel width and noise level, which are gradually increasing in the corresponding range [0.2, 2] and [5, 50] from left to right. We compare our UDVD to the most related work, SRMD~\cite{SRMD}, which handles the spatial variations without dynamic convolutions. As in Table~\ref{spatially}, UDVD constantly outperforms SRMD for all the settings on all the dataset, Set5~\cite{Set5}, Set14~\cite{Set14} and BSD100~\cite{BSD100}. In summary, UDVD delivers a noticeable PSNR improvement over SRMD. The qualitative comparison is also illustrated in Fig.~\ref{SVimg}. In Fig.~\ref{SVimg}, without considering spatial variations, RCAN can only handle a fixed degradation (red block represents lighter degradation), but fails at the other one (blue block represents heavier degradation). Although both SRMD and UDVD are capable to deal with spatial variations, UDVD still produces sharper and clearer reconstructed image. These results demonstrate that UDVD is capable to handle not only multiple degradations but also their spatial variations.

\vspace{-10pt}
\paragraph{Noise-Free Degradations and Variations.} In this experiment, we train UDVD for noise-free degradations (remove noise degradations) to enable extensive comparisons to other works. Table~\ref{noisefree} summarizes the PSNR results for noise-free UDVD and the competitive methods.Compare to SRMDNF~\cite{SRMD} and SFTMD ~\cite{SFTMD}, UDVD achieves comparable results in most cases and outperforms both the methods when large scale factor on BSD100~\cite{BSD100}. Note that SFTMD~\cite{SFTMD} benefits mostly from the Spatial Feature Transform (SFT) which applies an affine transformation on the degradation information rather than concatenating it with input image. Without losing generality, UDVD can also be extended to adopt SFT layers for further improvements.

\begin{table*}[t]
\small
\centering
\begin{tabular}{lcccccc}
\hline
\multirow{2}{*}{Methods} & \multicolumn{2}{c}{Set5}                                      & \multicolumn{2}{c}{Set14}                                      & \multicolumn{2}{c}{BSD100}                                     \\
             & \textbf{BI}                   & \textbf{DN}                   & \textbf{BI}                   & \textbf{DN}                   & \textbf{BI}                   & \textbf{DN}                   \\ \hline\hline
Bicubic         & 30.39/0.8308                   & 24.01/0.5369                   & 27.55/0.7271                   & 22.87/0.4724                   & 27.21/0.6918                   & 22.92/0.4449                   \\
SRCNN~\cite{SRCNN}    & 32.75/0.8944                   & 25.01/0.6950                   & 29.30/0.8074                   & 23.78/0.5898                   & 28.41/0.7736                   & 23.76/0.5538                   \\
VDSR~\cite{VDSR}     & 33.67/0.9150                   & 25.20/0.7183                   & 29.78/0.8244                   & 24.00/0.6112                   & 28.83/0.7893                   & 24.00/0.5749                   \\
SRMD$^*$~\cite{SRMD}     & 33.86/0.9232                   & 28.30/0.8123                   & 29.80/0.8342                   & 26.34/0.7025                   & 28.87/0.8001                   & 25.84/0.6561                   \\
RDN~\cite{RDN}      & \textcolor{blue}{34.71}/\textcolor{blue}{0.9280} & \textcolor{blue}{28.47}/\textcolor{blue}{0.8151} & \textcolor{blue}{30.57}/\textcolor{blue}{0.8447} & \textcolor{blue}{26.60}/\textcolor{blue}{0.7107} & \textcolor{blue}{29.26}/\textcolor{blue}{0.8079} & \textcolor{blue}{25.93}/\textcolor{blue}{0.6573} \\
RCAN~\cite{RCAN}     & \textcolor{red}{34.74}/\textcolor{red}{0.9299}  & -                        & \textcolor{red}{30.65}/\textcolor{red}{0.8482}  & -                        & \textcolor{red}{29.32}/\textcolor{red}{0.8111}  & -                        \\
UDVD$^*$          & 33.99/0.9240                   & \textcolor{red}{28.52}/\textcolor{red}{0.8186}  & 30.04/0.8371                   & \textcolor{red}{26.65}/\textcolor{red}{0.7146}  & 28.94/0.8016                   & \textcolor{red}{25.99}/\textcolor{red}{0.6632}  \\ \hline
\end{tabular}
\vspace{2pt}
\caption{Average PSNR/SSIM values on fixed degradations. “$*$” indicates a unified model for \textbf{BI} and \textbf{DN}. We use the provided official code to compute the results, except SRCNN and VDSR. For SRCNN and VDSR, results are extracted from the publications~\cite{SRCNN, VDSR, RDN}. The best two results are highlighted in \textcolor{red}{red} and \textcolor{blue}{blue} colors.}
\label{fixed}
\vspace{-8pt}
\end{table*}

\begin{figure*}[t]
\centering
\footnotesize
\begin{minipage}{0.19\textwidth}
\centerline{\includegraphics[width=\textwidth]{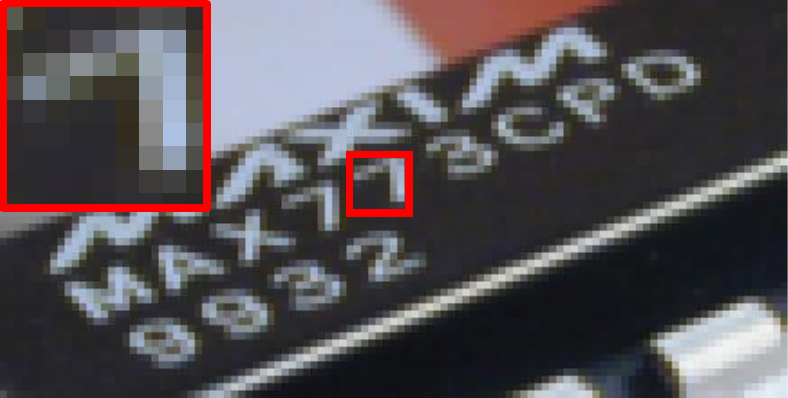}}
\centerline{\includegraphics[width=\textwidth]{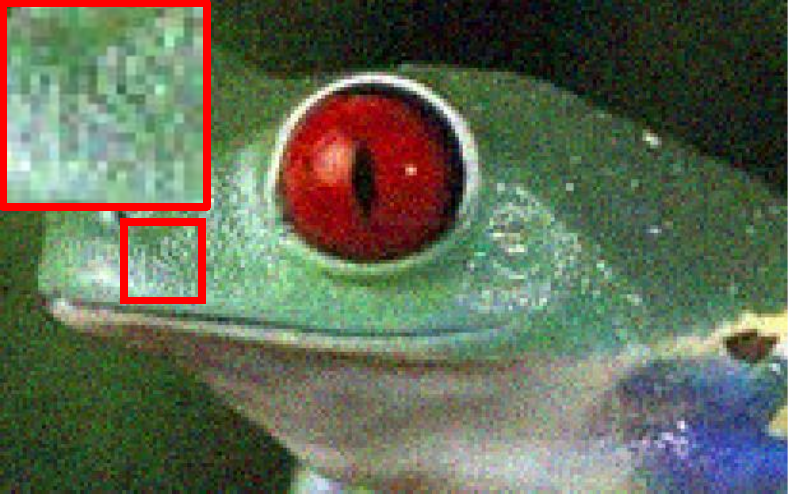}}
\centerline{(a) Real LR}
\end{minipage}
\begin{minipage}{0.19\textwidth}
\centerline{\includegraphics[width=\textwidth]{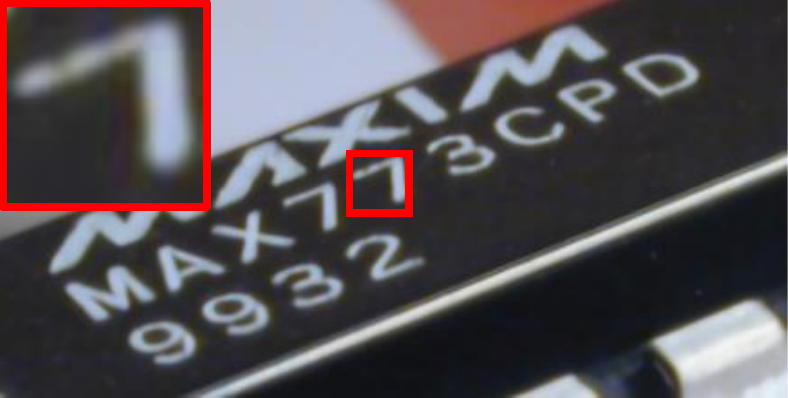}}
\centerline{\includegraphics[width=\textwidth]{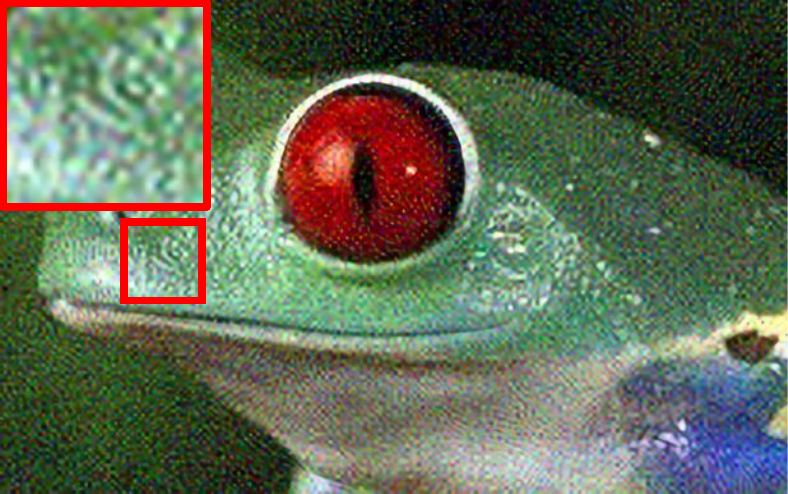}}
\centerline{(b) RCAN~\cite{RCAN}}
\end{minipage}
\begin{minipage}{0.19\textwidth}
\centerline{\includegraphics[width=\textwidth]{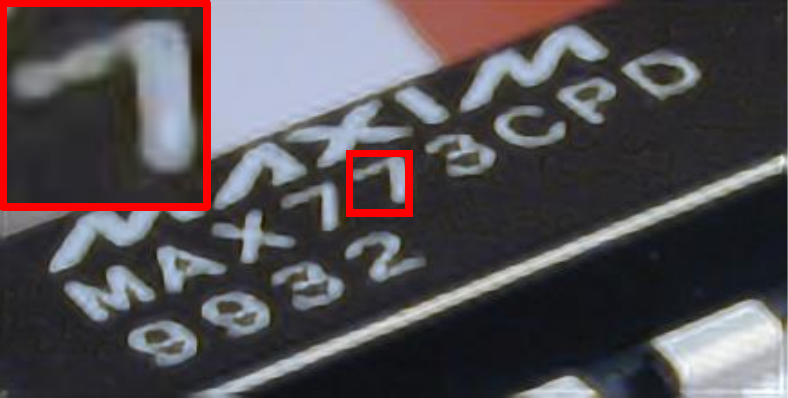}}
\centerline{\includegraphics[width=\textwidth]{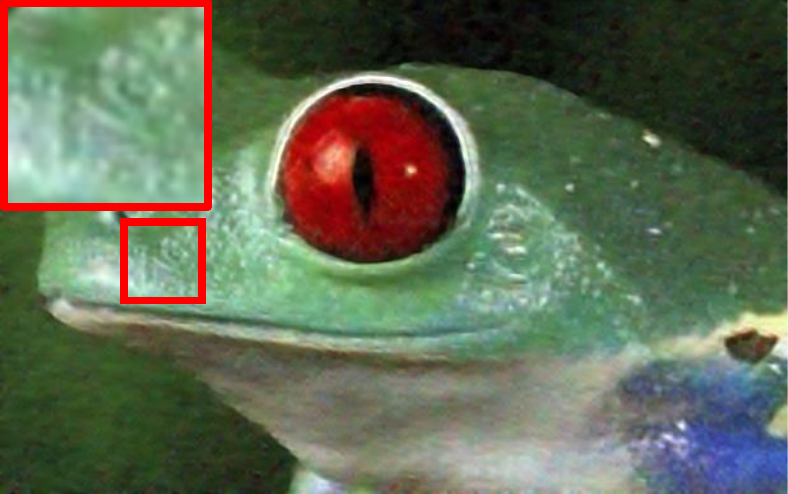}}
\centerline{(c) ZSSR~\cite{ZSSR}}
\end{minipage}
\begin{minipage}{0.19\textwidth}
\centerline{\includegraphics[width=\textwidth]{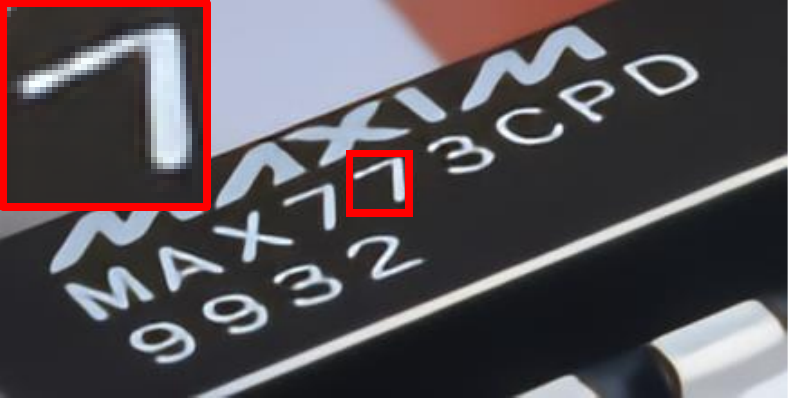}}
\centerline{\includegraphics[width=\textwidth]{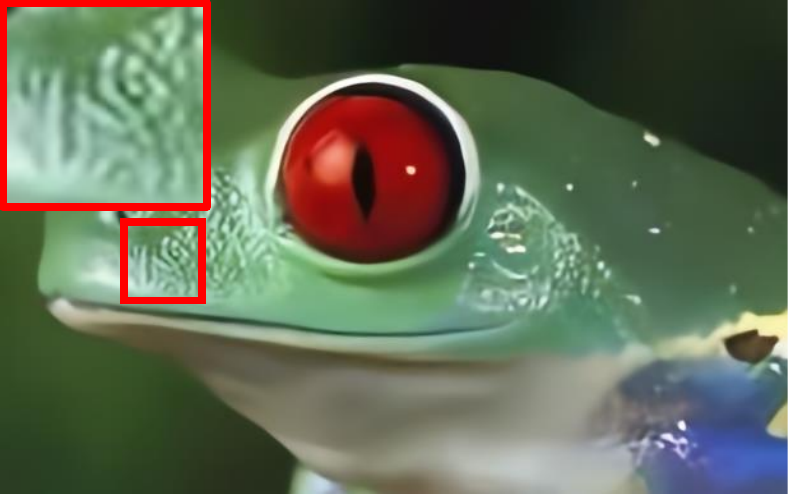}}
\centerline{(d) SRMD~\cite{SRMD}}
\end{minipage}
\begin{minipage}{0.19\textwidth}
\centerline{\includegraphics[width=\textwidth]{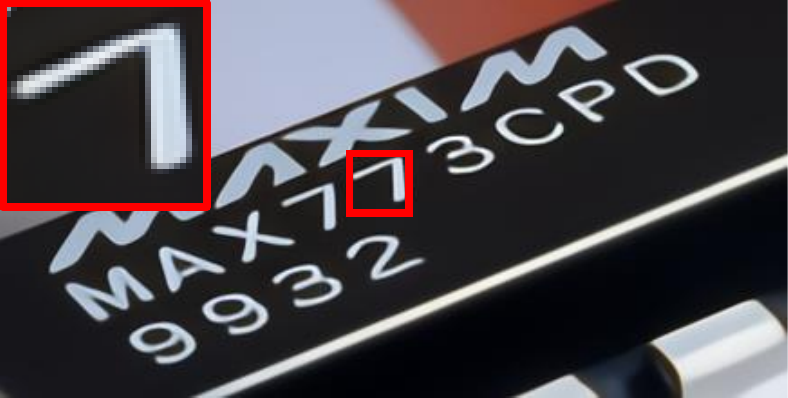}}
\centerline{\includegraphics[width=\textwidth]{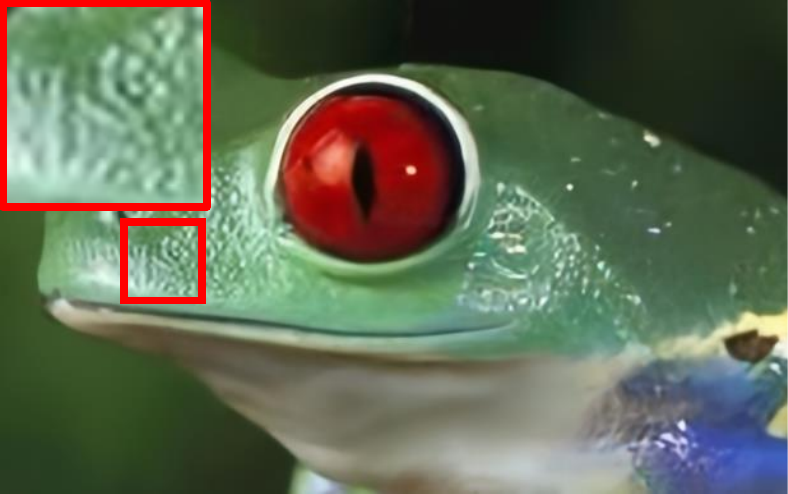}}
\centerline{(e) UDVD(Ours)}
\end{minipage}
\vspace{3pt}
\caption{The qualitative results of real image \textit{“chip”} with scale factor 4 (first row) and \textit{“frog”} with scale factor 3 (second row).}
\label{realimg}
\vspace{-12pt}
\end{figure*}

\begin{figure}
\centering
\footnotesize
\boldmath
\begin{minipage}{0.5\textwidth}
\centering
\begin{minipage}{0.4\textwidth}
\centerline{\includegraphics[width=\textwidth]{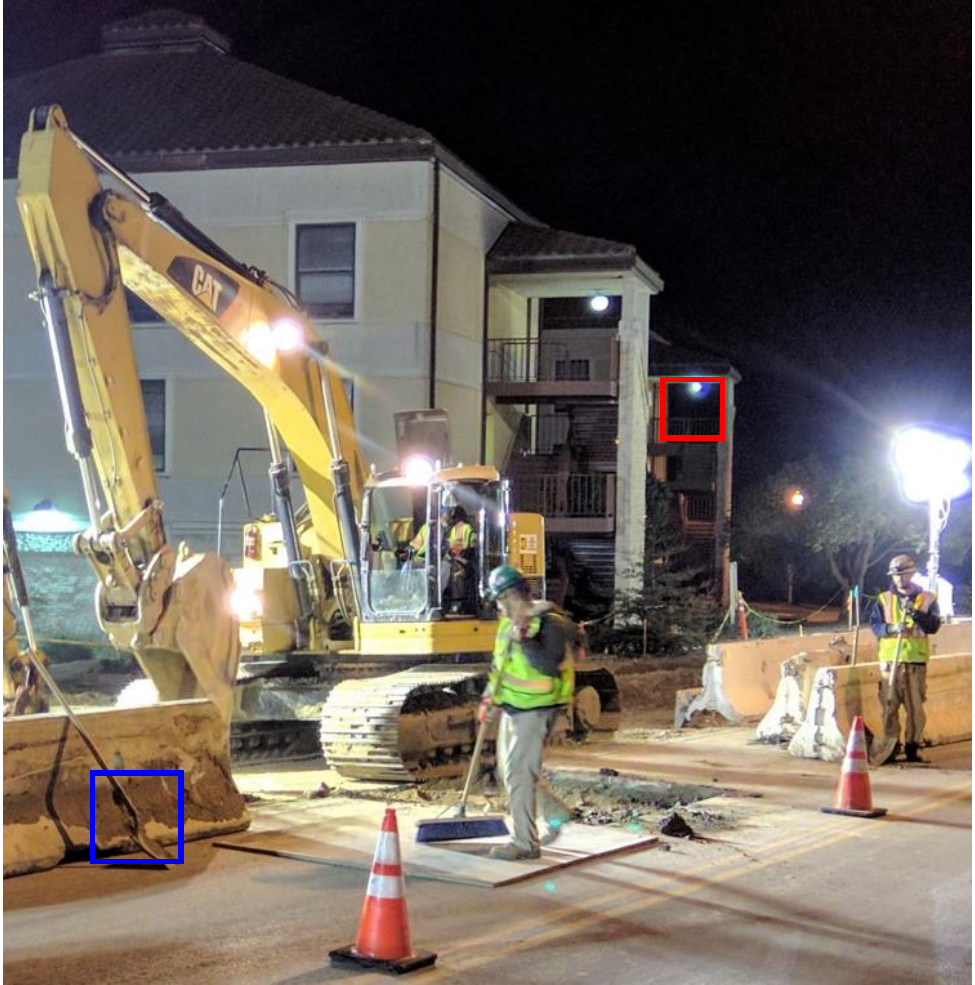}}
\vspace{-2pt}
\centerline{Real Image}
\end{minipage}
\begin{minipage}{0.2\textwidth}
\centerline{\includegraphics[width=\textwidth]{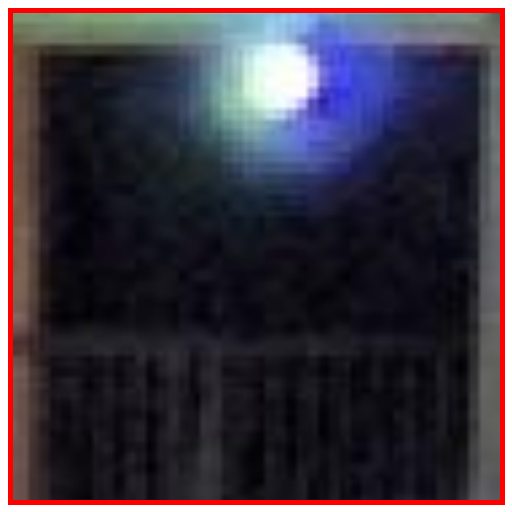}}
\centerline{\includegraphics[width=\textwidth]{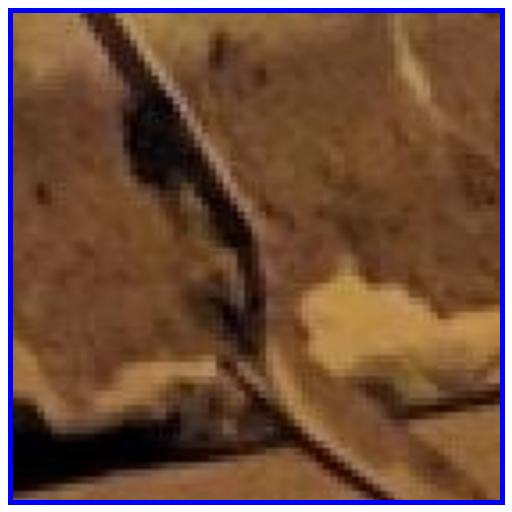}}
\vspace{-2pt}
\centerline{LR}
\end{minipage}
\end{minipage}

\begin{minipage}{0.5\textwidth}
\centering
\begin{minipage}{0.2\textwidth}
\centerline{\includegraphics[width=\textwidth]{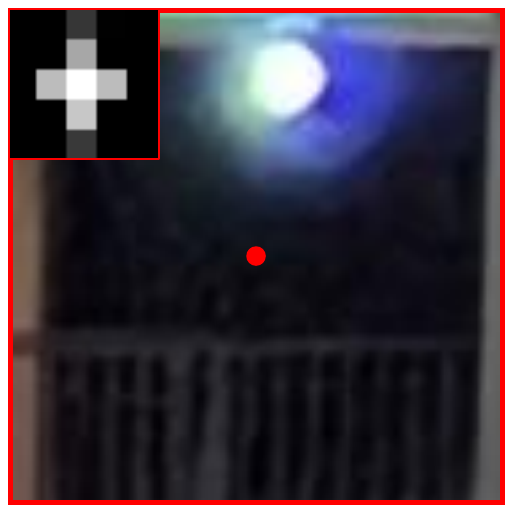}}
\centerline{\includegraphics[width=\textwidth]{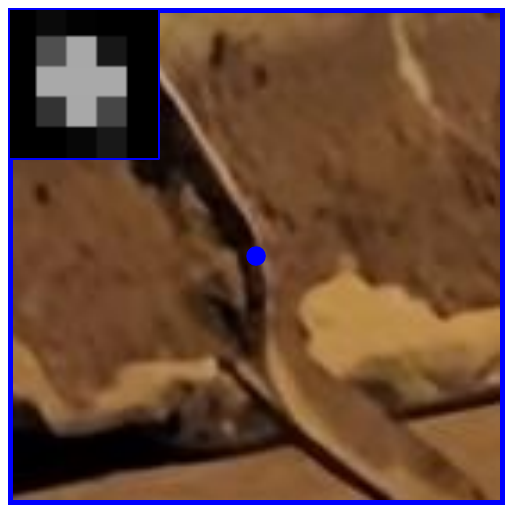}}
\vspace{-2pt}
\centerline{\textbf{$\varepsilon$=0.6, $\sigma$=5}}
\end{minipage}
\begin{minipage}{0.2\textwidth}
\centerline{\includegraphics[width=\textwidth]{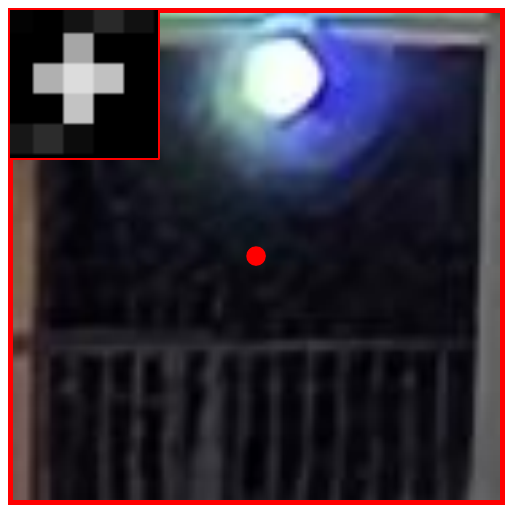}}
\centerline{\includegraphics[width=\textwidth]{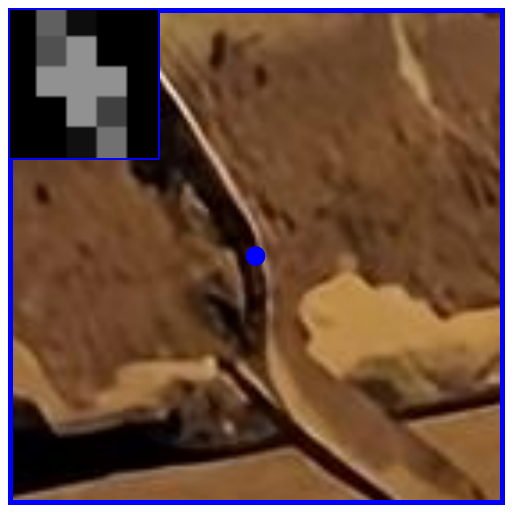}}
\vspace{-2pt}
\centerline{\textbf{$\varepsilon$=1.0, $\sigma$=5}}
\end{minipage}
\begin{minipage}{0.2\textwidth}
\centerline{\includegraphics[width=\textwidth]{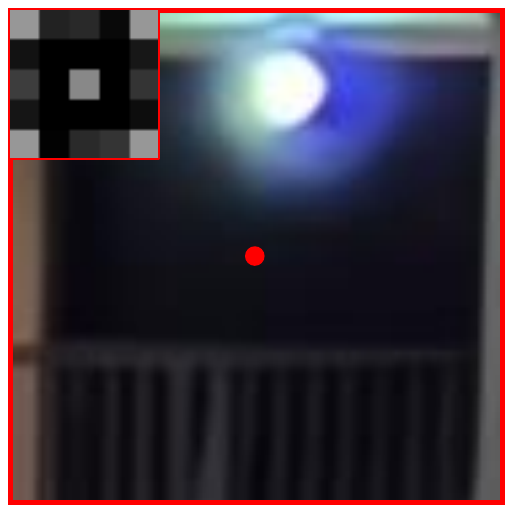}}
\centerline{\includegraphics[width=\textwidth]{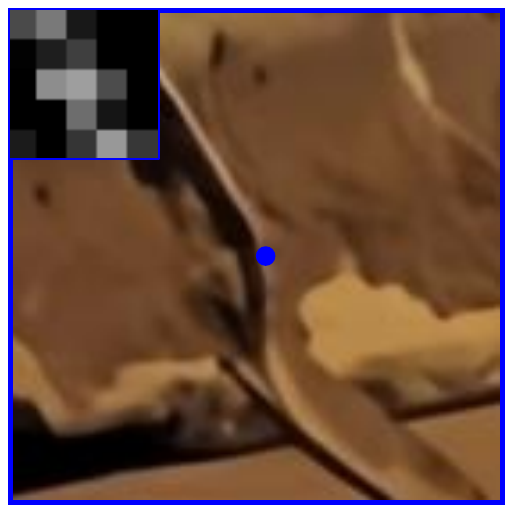}}
\vspace{-2pt}
\centerline{\textbf{$\varepsilon$=0.6, $\sigma$=10}}
\end{minipage}
\begin{minipage}{0.2\textwidth}
\centerline{\includegraphics[width=\textwidth]{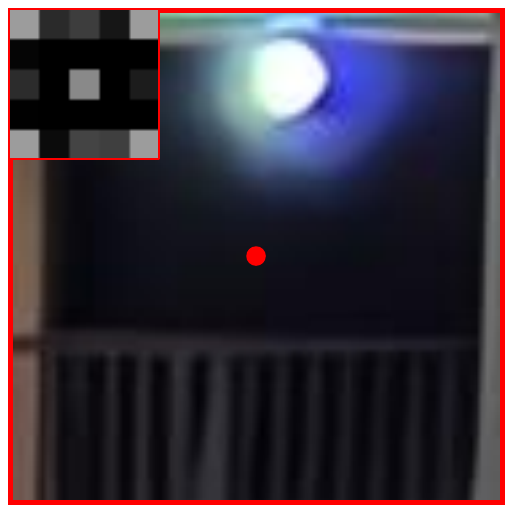}}
\centerline{\includegraphics[width=\textwidth]{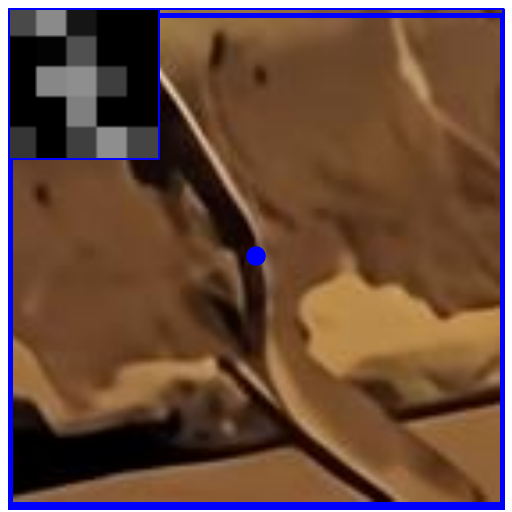}}
\vspace{-2pt}
\centerline{\textbf{$\varepsilon$=1.0, $\sigma$=10}}
\end{minipage}
\end{minipage}
\vspace{-2pt}
\caption{The qualitative results of real image with scale factor 2, variant Gaussian blur kernel width $\varepsilon$ and noise level $\sigma$.}
\vspace{-15pt}
\label{realspatial}
\end{figure}

\vspace{-10pt}
\paragraph{Fixed Degradations.} To compare with state-of-the-art fixed degradation based methods, we evaluate UDVD on two widely used fixed degradations \textbf{BI} and \textbf{DN}. \textbf{BI} only includes bicubicly downsampling. While \textbf{DN} applies bicubicly downsample and then add AWGN with noise level 30. Table~\ref{fixed} lists PSNR and SSIM results on \textbf{BI} and \textbf{DN} with scale factor 3. In Table~\ref{fixed}, except SRMD~\cite{SRMD} and UDVD, all the rest methods train different and specific models for \textbf{BI} and \textbf{DN} respectively. While SRMD~\cite{SRMD} and UDVD exploits a single model to handle both \textbf{BI} and \textbf{DN} degradations. With only a single model, UDVD still produces competitive results on \textbf{BI} and superior results on \textbf{DN}. This verifies that a single UDVD can adapt to various degradations and achieve promising results. 

\subsection{Experiments on Real Images}
\label{Experiments on Real Images}
Besides the experiments of synthetic test images in Section~\ref{Experiments on Synthetic Images}, we further extend the experiments to real images. There is no ground-truth degradations for real images. Hence, manual grid search on degradation parameters is performed as in \cite{SRMD} to obtain visually satisfying results. 
Fig.~\ref{realimg} illustrates the qualitative comparisons on the widely used real image \textit{“chip”}~\cite{chip} and \textit{“frog”}~\cite{frog} for UDVD, RCAN~\cite{RCAN}, ZSSR~\cite{ZSSR} and SRMD~\cite{SRMD}. Most of the compared methods produce noticeable artifacts. RCAN produces blur edges and cannot deal with noise. ZSSR tends to produce over-smoothed results. Although SRMD successfully removes these artifacts, but it fails to recover sharp edges. In contrast, the proposed UDVD reconstructs a sharper and clearer quality.
Fig.~\ref{realspatial} illustrates images reconstructed by UDVD with different degradation estimations. One can observe that different patches ({\color{blue}blue} and {\color{red}red}) have favorable reconstructed results on different degradation estimations. This confirms the existence of variational degradations in real images. In summary, Fig.~\ref{realimg} and Fig.~\ref{realspatial} confirm the effectiveness of UDVD for real images.


\section{Conclusion}
\label{Conclusion}
In summary, this paper presents a Unified Dynamic Convolutional Network for Variational Degradations (UDVD). We further introduce two types of dynamic convolutions to improve performance. Multistage loss is also applied to gradually refine images throughout the consecutive dynamic convolutions. With only a single network, the proposed UDVD efficiently handles a wide range of degradation variations for real-world images, including cross-image and spatial variations. UDVD is evaluated on both synthetic and real images with diverse variations of degrading effects. Comprehensive experiments are conducted to compare UDVD with various existing works. Through qualitative results, we confirm the effectiveness of dynamic convolutions over various existing works. Extensive experiments show that the proposed UDVD achieves favorable or comparable performance on both synthetic and real images.

{
\small\small
\balance
\bibliographystyle{ieee}
\let\oldbibliography\thebibliography
\renewcommand{\thebibliography}[1]{%
  \oldbibliography{#1}%
  \setlength{\itemsep}{0pt}%
}

\end{document}